\documentclass[11pt]{article}
\usepackage{amsmath,amsbsy,amsfonts,amssymb,graphics,epsfig,float,mathrsfs,amsthm,braket}
\usepackage{ifsym}
\usepackage{psfrag}
\usepackage{color}
\usepackage{caption}
\usepackage{subcaption}
\usepackage{multirow}
\usepackage{float}
\usepackage[normalem]{ulem}

\newcommand \beq{\begin{equation}}
\newcommand \eeq{\end{equation}}

\newcommand{\upd}{\mathrm{d}}
\newcommand{\Lop}{{\cal L}}

\newcommand{\Srr}{\Sigma_{rr}}
\newcommand{\Srq}{\Sigma_{r\theta}}
\newcommand{\wo}{\Delta}
\newcommand{\cF}{{\cal F}}

\newcommand{\s}[1]{{\textsf{\textbf{#1}}}}

\begin{document}
\title{\s{Static bistability of spherical caps}}
\author{ \textsf{Matteo Taffetani$^{1}$, Xin Jiang$^{2}$, Douglas P. Holmes$^{2}$ and Dominic Vella$^{1}$}\\ 
{\it$^{1}$Mathematical Institute, University of Oxford, UK}\\
{\it$^{2}$Department of Mechanical Engineering, Boston University, USA}}

\date{\today}
\maketitle
\hrule\vskip 6pt
\begin{abstract}
Depending on its geometry, a spherical shell may exist in one of two stable states without the application of any external force: there are two `self-equilibrated' states, one natural and the other inside out (or `everted'). Though this is familiar from everyday life -- an umbrella is remarkably stable, yet a contact lens can be easily turned inside out -- the precise shell geometries for which bistability is possible are not known. Here, we use experiments and finite element simulations to determine the  threshold between bistability and monostability for shells of different solid angle. We compare these results with the prediction from shallow shell theory, showing that, when appropriately modified, this offers a very good account of bistability even for relatively deep shells. We then investigate the robustness of this bistability against pointwise indentation. We find that indentation provides a continuous route for transition between the two states for shells whose geometry makes them close to the threshold. However, for thinner shells, indentation leads to asymmetrical buckling before snap-through, while also making these shells  more `robust' to snap-through. Our work sheds new light on the robustness of the `mirror buckling' symmetry of spherical shell caps.
\end{abstract}
\vskip 6pt
\hrule

\maketitle

\section{Introduction}
Stability is a crucial aspect in the design and  construction of engineering structures; understanding their stability has therefore been the primary motivation for much classic analysis in structural mechanics. More recently, structures that are able to switch between two different configurations have attracted interest for applications in which morphing between states is  desirable, for example  in mechanical metamaterials and origami structures \cite{Andres2014_langmuir,Daynes2014_sms, Kamrava2017_sr} or in actuators \cite{Arrieta2011_jimss}. Various mechanisms have been proposed by which the system may be forced between two stable states  including loading via magnetic forces \cite{Loukaides2014,Seffen2016_sms}, fluid flow  \cite{Arena2017,Gomez2017b, Tavakol2016}, changes in lateral confinement \cite{Holmes2007,Gomez2017a} or, more generally, variations in the natural curvature of a structure \cite{Pezzulla2017_prsa, Pezzulla2018}.

\begin{figure}
\centering
\begin{subfigure}{0.45 \linewidth} 
	\includegraphics[width=1\linewidth]{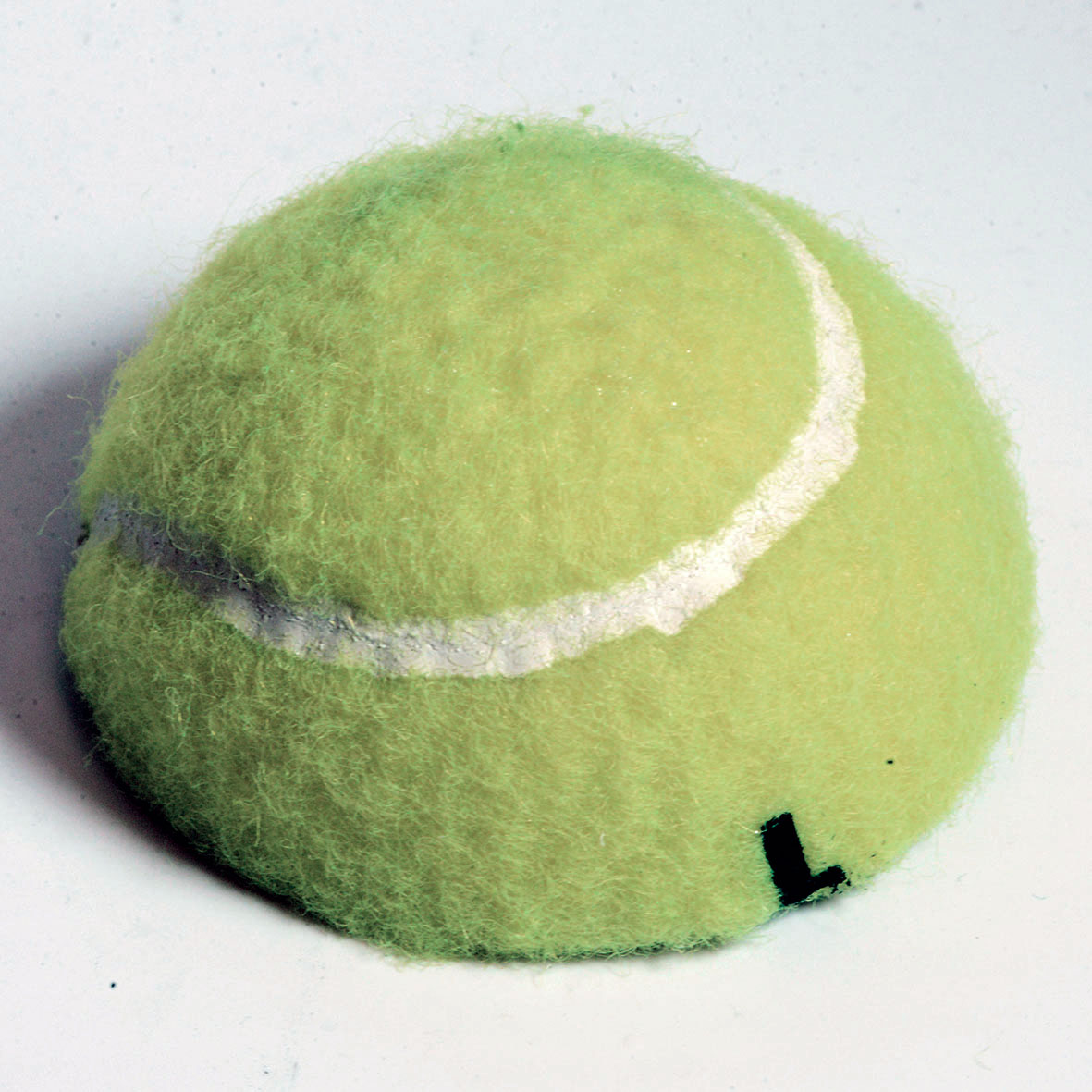}
	\caption{}
\end{subfigure}
\begin{subfigure}{0.45 \linewidth}
	\includegraphics[width=1\linewidth]{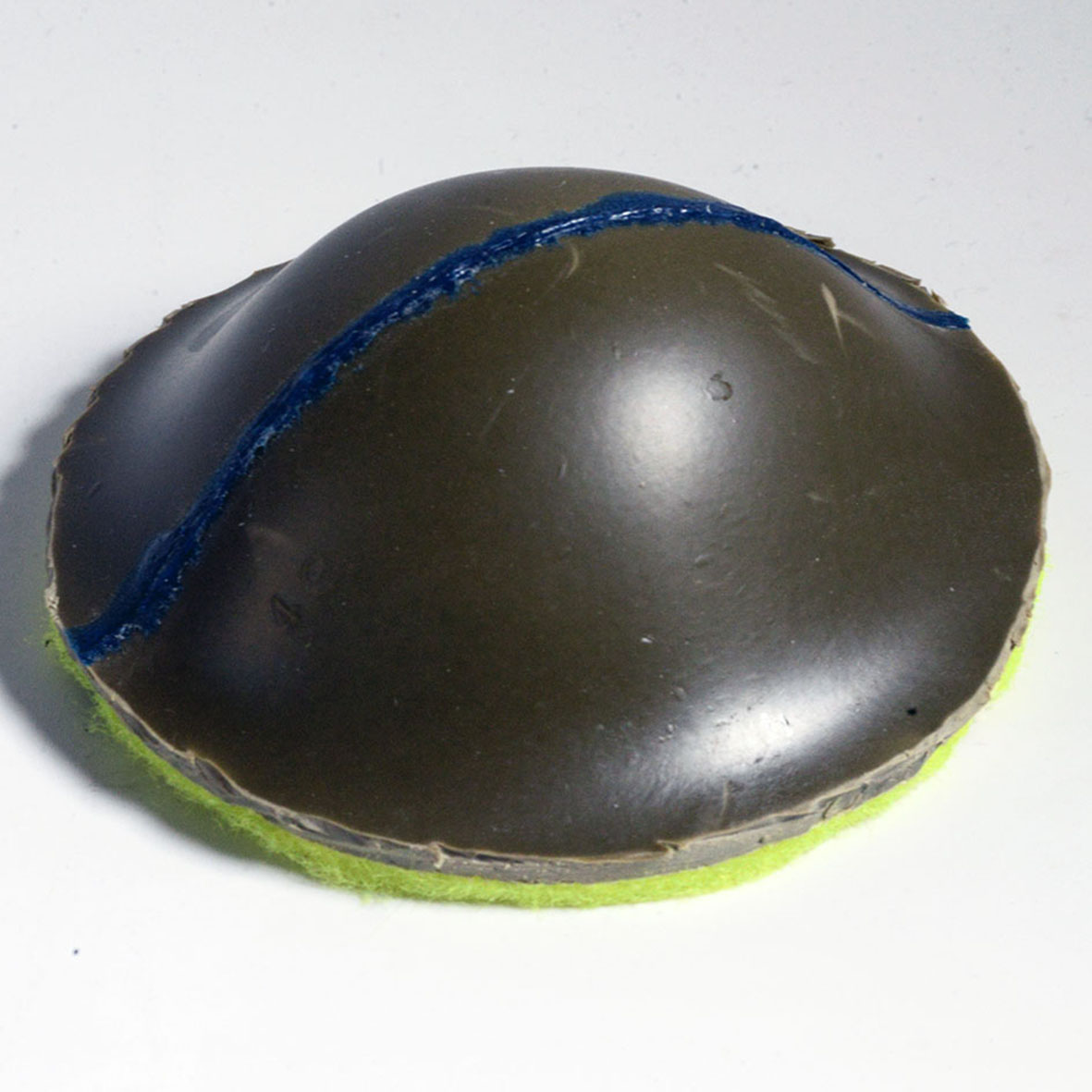}
	\caption{}
\end{subfigure}
\caption{The bistability of a spherical cap. (a) A section of a tennis ball in its natural state (before being everted). (b) The same section of a tennis ball as in (a) but now everted.} \label{FIG:shell_geometry}
\end{figure}

Perhaps the simplest example of bistability is exhibited by spherical cap shells, and can be demonstrated by cutting a section of a tennis ball (see figure \ref{FIG:shell_geometry}a,b). Provided that the section of the shell is sufficiently `deep', it may be turned `inside out', or everted, and it will remain in this everted state when the loading is removed. This bistability is related to the fact that an infinitely thin shell can be turned inside out to make a spherical cap of the same radius without stressing the centreline of the shell --- eversion represents an isometry of the shell, known as `mirror buckling', that does not cost any elastic energy in the limit of infinitely thin shells \cite{Pogorelov,Gomez2016_prs}. While the everted state is a good approximation for very thin shells, a shell of finite thickness requires a small bending moment to be applied at  edges to retain the spherical shape. In the absence of such a moment being applied, an everted shell will typically bend slightly near the edge, see  figure \ref{FIG:shell_geometry}b; in this edge region the shell is `trying' to return to its natural state. At a heuristic level, for caps that are too `shallow', this bending boundary layer \cite{LibaiSimmonds} extends across a significant portion of the shell and renders the everted state unstable: the shell immediately returns to its natural state once the applied loads are removed and the cap is  monostable. 

When the shell is monostable, but close to the transition, the dynamics of the snap-through slows down so that the cap is sometimes described as being `pseudo-bistable'\cite{Brinkmeyer2012_ijss}. While the reasons for this slow dynamics remain an area of active research \cite{Gomez2017a, Efrati2017}, it is a little surprising that previous studies have only skirted around the question of when a spherical shell cap is bistable --- indeed, Libai \& Simmonds \cite{LibaiSimmonds} are only able to offer rough estimates of the transition based on inferences from previous studies \cite{Mescall1965_jam,Brodland1987_ijss}. More recently, Brinkmayer {\it et al.} \cite{Brinkmeyer2012_ijss} used finite element simulations to determine the condition for bistability in deep spherical shells, focusing only on nearly incompressible shells.

Some insight into the transition from monostability to bistability has been given by various approximate analytical techniques. For example, Seffen \cite{Seffen2007_prsa} investigated the bistable configurations of orthotropic shallow shells with elliptical planforms (with and without twist) by assuming that both the undeformed and the deformed configurations maintain constant curvatures. Vidoli \cite{Vidoli2013_ijss} extended this approach by considering, in addition to the constant curvature case, various \emph{ans\"{a}tze} for the normal displacements that allow for linearly varying curvature and quadratically varying curvature in shells of varying planform. (This last condition was associated with the requirement that the boundary condition of zero bending moment be satisfied only on average.) Some of the results of this work were subsequently employed by Seffen and Vidoli \cite{Seffen2016_sms} to investigate the eversion of bistable shells under magnetic actuation. While these analytical approaches do allow for some progress, they have two main limitations: (i) to identify the threshold for bistability they employ shallow shell theory (which might seem to be an inappropriate simplification given that everyday experience suggests that the transition to bistability is associated with the depth of the shell) and  (ii) even the simplified shallow shell theory is highly nonlinear, so that analytical progress is only possible in certain limits. Furthermore, simplified models of deformation, such as mirror buckling\cite{Pogorelov}, may introduce substantial quantitative errors, despite appearing qualitatively correct \cite{Gomez2016_prs}.

At the simplest level, therefore, the question remains of when bistability occurs (in the absence of applied loads); a subsidiary question is whether shallow shell theory is able to give a good account of this transition. However, the relationship between mirror buckling and the robustness of bistability also poses some questions. For example, for shells that are only just deep enough to be bistable, we expect that the perturbation from gently touching the everted state may be enough to cause them to snap to the natural state. Even for  shells that are deep enough to be bistable there appear to be some complications: it is expected that indentation by a large amount $\delta$ should lead to mirror buckling over a horizontal length scale $(\delta R)^{1/2}$ where $R$ is the shell's radius of curvature\cite{Pogorelov,Gomez2016_prs}. However, it is also known that  a spherical shell that is indented does not in practice approach the mirror buckled solution. Instead, several studies have shown that such an indented spherical cap shell will form a polygonal buckling pattern at a critical indentation depth \cite{Fitch1968_ijss,Vaziri2008_pnas}. Nevertheless, if indentation proceeds through this transition, axisymmetry is at some point regained, if only when the shell eventually snaps through to the everted state. The question is then why/how does mirror buckling re-emerge at very large deformations close to total eversion?

\begin{figure}[H]
\centering
\begin{subfigure}{0.35 \linewidth} 
	\includegraphics[width=1\linewidth]{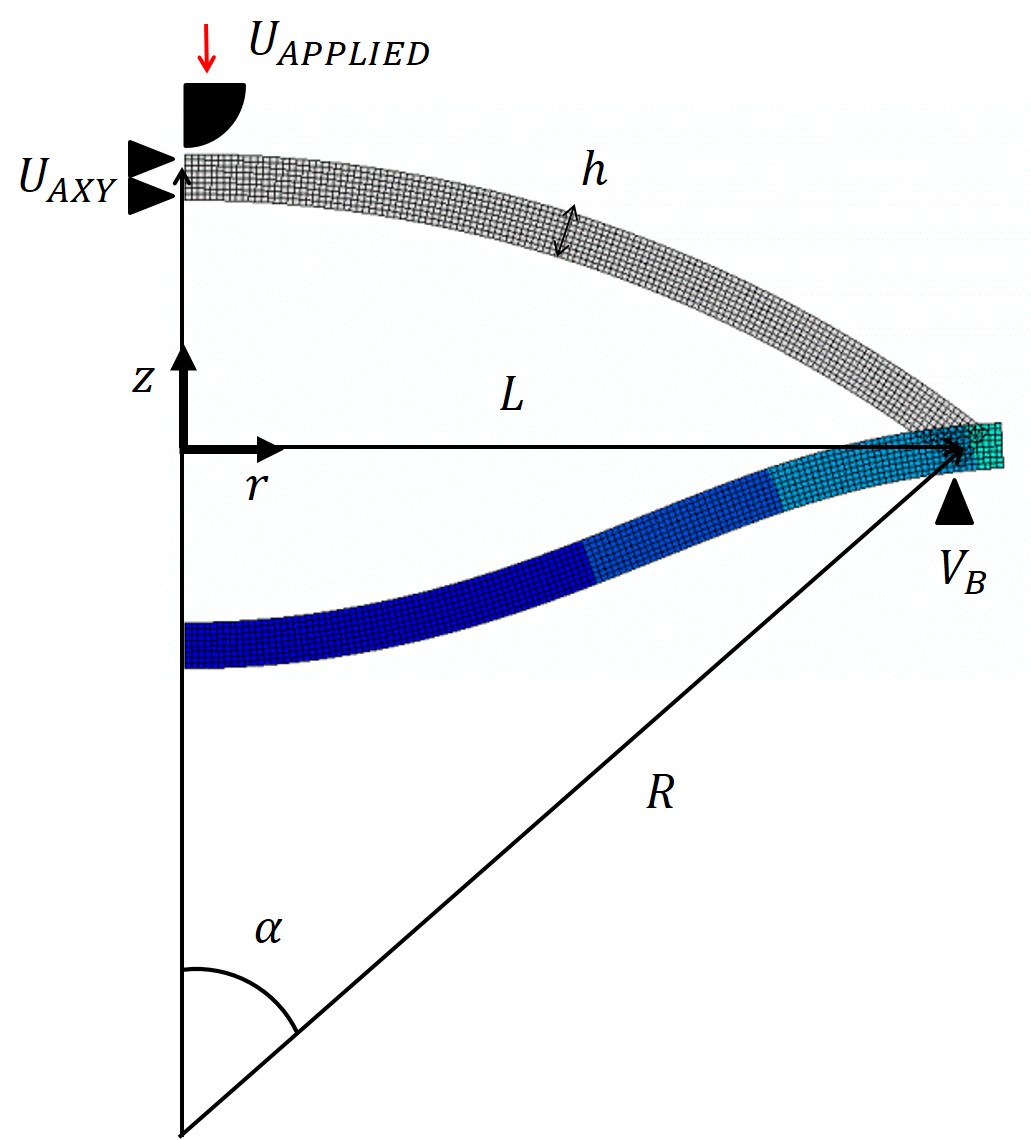}
	\caption{}
\end{subfigure}
\begin{subfigure}{0.6 \linewidth}
	\includegraphics[width=1\linewidth]{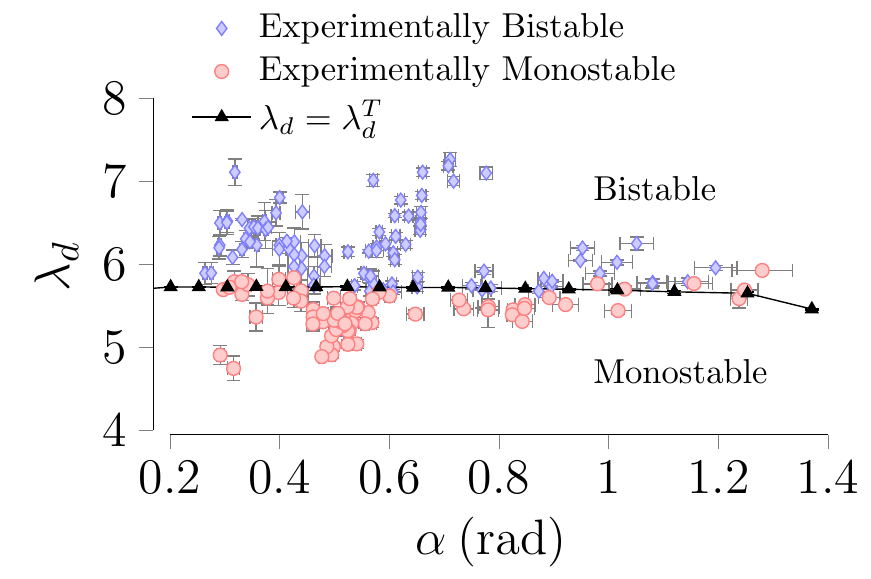}
	\caption{}
\end{subfigure}
\caption{(a) Parameters in our ABAQUS model (b) The $(\alpha,\lambda_d)$ parameter space, showing regions for which a shell is bistable or monostable. ABAQUS simulations predict that the shell should be monostable when $\lambda_d<\lambda_d^T(\alpha)$ (continuous line and triangles), and bistable for $\lambda_d>\lambda_d^T(\alpha)$, with $\lambda_d$ defined in equation \eqref{EQ:lambda_d}. This prediction is consistent with experiments on PVS, $\nu\approx0.5$, with points representing shells that were bistable (blue diamonds) or monostable (red circles).}
\label{FIG:lambda_experiments}       
\end{figure}

In this paper we focus on the bistability and buckling of spherical shell caps. We begin by investigating the bistability of axisymmetric spherical shells with a free external boundary and evaluating the ability of shallow shell theory to accurately predict the transition from bistability to monostability. In Section \ref{SEC:Bistability} we use experimental tests and finite element simulations to precisely determine the threshold between monostability and bistability as a function of the geometrical  parameters of the shell. In particular, we explore how this threshold varies as the depth of the shell (measured by the angular opening of the shell, $\alpha$, defined in figure \ref{FIG:lambda_experiments}a); we find that the predictions of shallow shell theory, suitably rescaled, hold even as $\alpha\to\pi/2$. We then study, in Section \ref{SEC:Indentation}, the robustness of the bistable state by considering how indentation modifies the stress state of the shell. In doing so, we present a  regime diagram for the behaviours of a spherical shell cap, which shows when the shell is bistable or monostable, and when it is susceptible to polygonal buckling upon indentation. We summarize our findings in \S\ref{SEC:Conclusion}.

\section{Bistability of free axisymmetric shells}\label{SEC:Bistability}

Our first goal is to understand the geometrical conditions under which a deep spherical shell is bistable, i.e.~when does a configuration other than the \emph{natural} one exist without the application of any external loads? Since we are interested in the behaviour of a shell in a self-equilibrated state, the only quantity in the problem that has units of force (more properly pressure) is the Young modulus $E$. With no other force scale in the problem, dimensional considerations alone lead us to expect that the results should be independent of the Young modulus. Instead, the transition between bistability and monostability should depend on the geometrical properties of the shell, e.g.~its thickness, $h$, radius of curvature $R$ and, perhaps, Poisson's ratio $\nu$. However, a key further variable is the `depth' of the shell, which, loosely speaking represents the proportion of a complete sphere that is contained in the shell. A more precise measure of the `depth' of a shell is the solid angle $A=S/R^2$, where $S$ is the surface area of the shell: a sphere corresponds to $A=4\pi$, while a hemisphere has $A=2\pi$. In our axisymmetric framework $A$ can be easily related to the (planar) angle $\alpha$  subtended between the `pole' and the free edge  (see figure \ref{FIG:lambda_experiments}a) as $A=2 \pi \left(1- \cos \alpha\right)$. If this $\alpha\ll1$, the shell is shallow \cite{Ventsel2001}, but for $\alpha=O(1)$ the shell is deep. To understand further how this depth influences bistability, we consider the various energy scales that enter the problem. The relevant energies quantify the bending and stretching of the shell, and while a more formal analysis is possible, we proceed here with a heuristic discussion aimed at giving physical insight. The bending energy density induced by eversion is $\mathcal{E}_B \sim B \left(1/R\right)^2$, where $B=E h^3/[12(1-\nu^2)]$ is the bending stiffness of the shell, and arises from the compression of the outer surface and stretching of the inner surface. We estimate the stretching energy density by estimating the strain due to eversion, $\epsilon\sim \delta^2/\ell^2$ with $\ell\sim \sqrt{\delta R}$ the characteristic horizontal length associated with mirror buckling, as discussed above and by Pauchard \emph{et al.} \cite{Pauchard1998_PMb}, for example. Elementary geometry gives that the vertical deflection, $\delta=R(1-\cos\alpha)\sim \alpha^2R$, leading to $\epsilon\sim\alpha^2$ and hence to an estimate of the stretching energy density $\mathcal{E}_s\sim Eh \epsilon^2 \sim Eh\alpha^4$. 
The relative importance of stretching to bending energy densities is therefore
\begin{equation}
\frac{\mathcal{E}_s}{\mathcal{E}_b} \sim \frac{E h \alpha^4}{B/R^2} = 12 \left(1-\nu^2\right)\frac{R^2}{h^2}\alpha^4.
\end{equation}
This ratio measures the importance of bending and stretching energies, but the behaviour of shells is more conventionally studied in terms of the fourth root of this parameter \cite{Brinkmeyer2012_ijss, LibaiSimmonds}, namely
\begin{equation}\label{EQ:lambda_d}
\lambda_d = \left[12 \left(1-\nu^2\right)\right]^{1/4}\sqrt{\frac{R}{h}}\alpha.
\end{equation} We note that the parameter $\lambda_d$ involves both the depth of the shell, measured via $\alpha$, and its slenderness, $h/R$. While for shallow shells we expect $\alpha\ll1$, finite values of $\lambda_d$ may be obtained by taking small values of $h/R$. Our aim now is to characterize the bistability of the shell in terms of the two parameters $\lambda_d$ and $\alpha$. We do this using a combination of experiments, numerical simulations in ABAQUS, and shallow shell theory \cite{Ventsel2001}. \

Details of the experimental and simulation protocols are given in Appendix \ref{APP:MatMet}. In short, we fabricated shells of different geometrical properties using Polyvinylsiloxane (Zhermack) and tested whether these shells were bistable or not. In the ABAQUS model, we simulated indenting shells of different geometries to the everted state and observed whether the shell remained in this configuration when the loading was removed. Figure \ref{FIG:lambda_experiments} shows the behaviour (monostable or bistable) throughout the $(\alpha,\lambda_d)$ parameter space, as predicted from ABAQUS simulations of an incompressible material, $\nu=0.5$. For a given value of $\alpha$, the critical value of $\lambda_d$  at the transition, $\lambda_{d}^{T}$, is determined to within 2.5\%; this threshold value is identified as the smallest value of $\lambda_{d}$ for which bistability is observed. Figure \ref{FIG:lambda_experiments} also shows points corresponding to shells that were experimentally observed to be bistable (diamonds) or monostable (circles). 

We note two things from the regime diagram presented in figure \ref{FIG:lambda_experiments}: first, experiments are entirely consistent with the finite element predictions, since red circles (monostable) cluster below the triangles (simulation results) while blue diamonds (bistable) cluster above the triangles. Second, and more remarkably, we note that the monstable/bistable threshold does not significantly vary with the shell depth $\alpha$, except very close to the hemispherical case, $\alpha=\pi/2$. This numerical and experimental observation suggests that a simpler, and perhaps more insightful, approach to the problem is to consider the shallow shell limit, $\alpha\to0$. In this case, the governing equations simplify somewhat, becoming the Donnell--Mushtari--Vlasov equations \cite{Ventsel2001}. This allows a little more insight into the mechanics governing eversion, and also simplifies the numerical calculations required to determine the threshold between monostability and bistability, $\lambda_d^T$.

\subsection{Shallow shells}

When the angle $\alpha$ (see figure \ref{FIG:lambda_experiments}a) is small, the shell is \emph{shallow}: the lateral extent of the shell is such that $L\ll R$. The simplification that $\alpha\ll1$ allows us to write $\alpha\approx L/R$ and hence the appropriate dimensionless parameter $\lambda_d$ is replaced by its shallow version,
\begin{equation}\label{EQ:lambda_s}
\lambda_s = \left[12 \left(1-\nu^2\right)\right]^{1/4}\frac{L}{(hR)^{1/2}}.
\end{equation} This parameter may now be interpreted as the lateral size of the shell, $L$, compared to the width of the typical boundary layer that is induced by the competition between bending and stretching, $(hR)^{1/2}$. If the shell is large compared to this boundary layer, we expect it should remain bistable, but if the boundary layer becomes too large then it will instead be monostable, similar to the threshold for bistability in spherical caps subjected to an evolving natural curvature~\cite{Pezzulla2018}.

To proceed more formally, we note that in the shallow limit, the mechanics of a shell is close to that of its projection onto the plane, with additional terms capturing the leading order effect of the shell's curvature. These Donnell--Mushtari--Vlasov (DMV) equations \cite{Ventsel2001,Vella2011_prl} may be expressed in terms of a cylindrical coordinate system $(r, \theta, z)$ such that the apex of the shell lies along the $z$-axis and the shell's planform lies in the $(r, \theta)$ plane.

To obtain bistability within the DMV framework requires that the centerline of the shell may stretch, though we emphasize that in the theory presented here, strains remain small, and the material behaves in a Hookean manner throughout. To determine the normal displacement of the shell's centreline, $w(r,\theta)$, we use the normal force balance \cite{Vella2011_prl}
\beq \label{EQ:force_balance}
B\nabla^4 w + \frac{1}{R} \nabla^2 \phi - \Lop\left(\phi,w\right) = 0,
\eeq where the Airy stress function, $\phi(r,\theta)$, is introduced to ensure that the in-plane equilibrium equations are automatically satisfied \cite{Box2017_prsa}. However, to ensure that the strains corresponding to a given stress function $\phi(r,\theta)$ are compatible, requires an additional equation:
\beq\label{EQ:compatibility}
\frac{1}{Eh}\nabla^4 \phi + \tfrac{1}{2} \Lop\left(w,w\right) - \frac{1}{R}\nabla^2w =0.
\eeq 
Here $\nabla^2$ is the Laplacian operator in cylindrical coordinates and the operator $\Lop(\cdot,\cdot)$ is  defined \cite{Ventsel2001} by
\beq
\Lop\left(f,g\right)=\frac{\partial^2 f}{\partial r^2}\left(\frac{1}{r}\frac{\partial g}{\partial r} + \frac{1}{r^2} \frac{\partial^2 g}{\partial \theta^2}\right)+\frac{\partial^2 g}{\partial r^2}\left(\frac{1}{r}\frac{\partial f}{\partial r} + \frac{1}{r^2} \frac{\partial^2 f}{\partial \theta^2}\right) - 2 \frac{\partial}{\partial r}\left(\frac{1}{r} \frac{\partial f}{\partial \theta}\right) \frac{\partial}{\partial r}\left(\frac{1}{r} \frac{\partial g}{\partial \theta}\right).
\eeq

To non-dimensionalize the problem, we let
\begin{equation}
W = \frac{w R}{L^2}; \;\;\; \rho =\frac{r}{L}; \;\;\; \Phi =\frac{\phi R^2}{Y L^4},
\end{equation} so that the shallow shell equations \eqref{EQ:force_balance}--\eqref{EQ:compatibility} can be rewritten as
\beq\label{EQ:vertical_force_normalizedFULL}
\frac{1}{\lambda_s^4}\nabla^4 W + \nabla^2 \Phi - \Lop \left(\Phi, W\right) =0
\eeq
\beq\label{EQ:compatibility_normalizedFULL}
\nabla^4 \Phi + \tfrac{1}{2} \Lop\left(W,W\right) - \nabla^2 W=0.
\eeq In \eqref{EQ:vertical_force_normalizedFULL} we again see the appearance of the dimensionless parameter $\lambda_s$, defined in \eqref{EQ:lambda_s}, as the dimensionless system size.

The problem in Equation \eqref{EQ:vertical_force_normalizedFULL} and \eqref{EQ:compatibility_normalizedFULL} consists of two fourth-order partial differential equations, and so eight boundary conditions are required to close it. Details of the relevant boundary conditions for different problems are given explicitly in table \ref{TAB:BCs_zeroth}, but here we limit ourselves to a qualitative description: we require symmetry and regularity at the apex, $\rho=0$, while the shell is simply supported along the outer boundary (dimensionless position $\rho=1$). In particular, the shell is free to rotate around the azimuthal direction, i.e.~the bending moment $M_r=0$. In cylindrical coordinates, this condition may be written:
\begin{equation}\label{EQ:Mr}
M_r=\left.\left(\frac{\partial^2 W}{\partial \rho^2}+\frac{\nu}{\rho}\frac{\partial W}{\partial \rho}+\frac{\nu}{\rho^2}\frac{\partial^2 W}{\partial \theta^2}\right)\right|_{\rho=1}=0.
\end{equation} Other boundary conditions are expressed in table \ref{TAB:BCs_zeroth}; for the results presented in the main text we use the `free' boundary conditions listed there.

\begin{figure}
\centering
\begin{subfigure}{0.49 \linewidth}
	\includegraphics[width=0.9\linewidth]{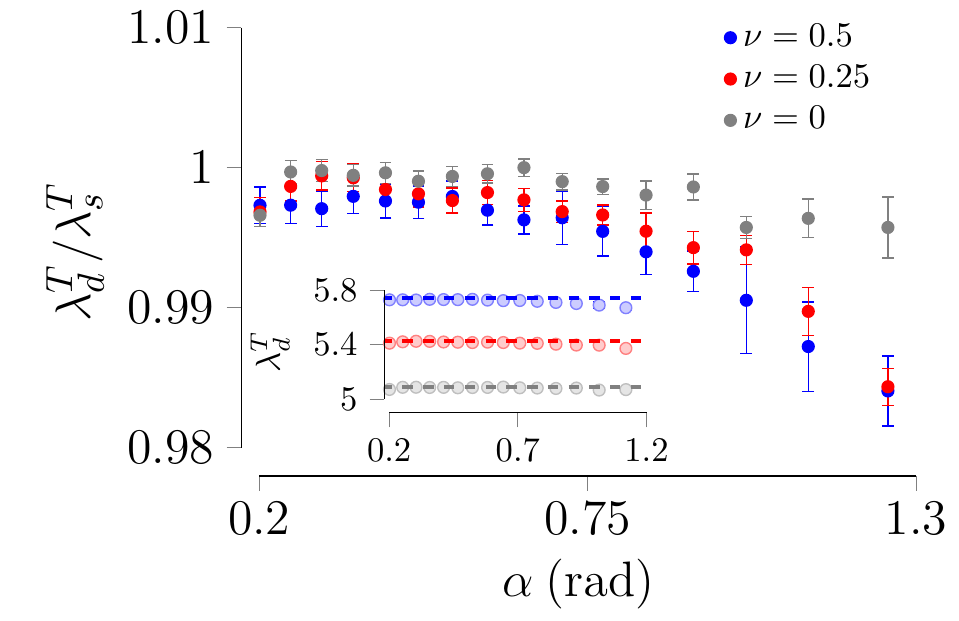}
	\caption{}
\end{subfigure}
\begin{subfigure}{0.49 \linewidth}
	\includegraphics[width=0.9\linewidth]{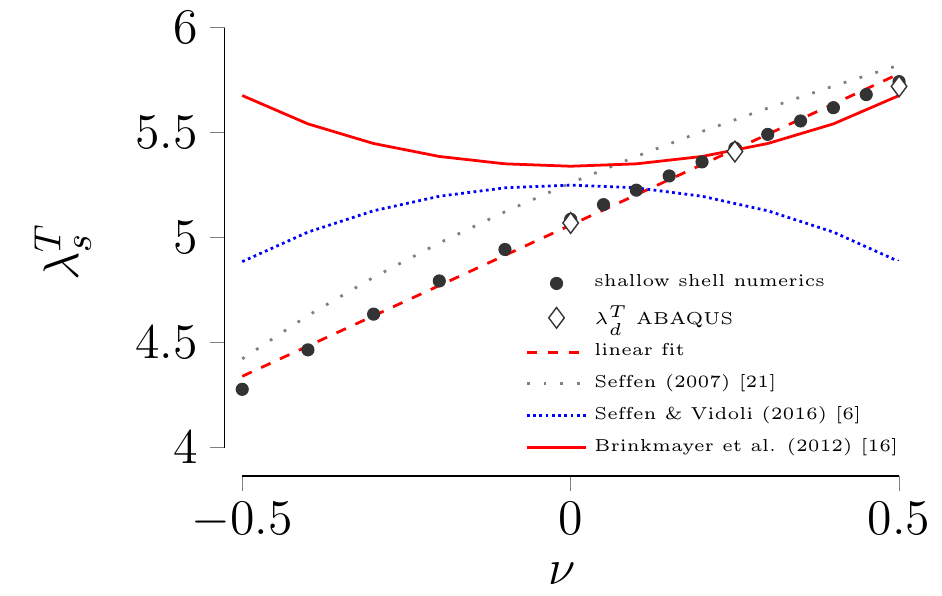}
	\caption{}
\end{subfigure}
\caption{(a) The threshold value of $\lambda_d$ for bistability, $\lambda_{d}^T$, determined using ABAQUS simulations for different sections of the shell ($\alpha$) and Poisson's ratio $\nu$. Results are normalized by the corresponding, numerically determined, threshold for shallow shells, $\lambda_s^T$; this shows that $\lambda_s^T$ gives a good account of $\lambda_d^T$ for all investigated $\alpha$, despite only being strictly valid as $\alpha\to0$. (b) The threshold for the transition from monostable to bistable, as predicted by shallow shell theory,  is plotted as a function of Poisson's ratio, $\nu$ (points).  The dashed red line indicates the  best fit line $\lambda_s^T=1.44\nu+5.06$. For comparison, other curves indicate  three previous proposals for this relationship: numerical correction proposed by Brinkmeyer et al.~\cite{Brinkmeyer2012_ijss} for the region $0.4\leq\nu\leq0.49$ and extended to the entire range (solid curve), semi-analytical results from ref.~\cite{Seffen2007_prsa} (dotted curve) and ref.~\cite{Seffen2016_sms} (narrow dotted curve), adapted to the present geometry; the three diamonds represent the results of our ABAQUS simulations with $\alpha=0.2$ and three values of $\nu$ shown in the inset of (a).}\label{FIG:lambda_normalized}
\end{figure}

In the complete problem \eqref{EQ:vertical_force_normalizedFULL}--\eqref{EQ:Mr}, there are now two dimensionless parameters: the dimensionless system size, $\lambda_s$, defined in \eqref{EQ:lambda_s}, and the Poisson's ratio $\nu$. (Note that $\lambda_s$ also depends on $\nu$.) Restricting ourselves to axisymmetric deformations,  we use the numerical continuation software AUTO-0p7p \cite{Doedel2012} to investigate the critical value of $\lambda_s$ at which the monostable/bistable threshold occurs for different values of the Poisson's ratio $\nu$ in the range $-0.5\leq\nu\leq0.5$.  In this numerical continuation analysis, our code starts from the isometric eversion {\em ansatz} with $\lambda_s=30$: this is a large value for which we expect this starting guess to represent a good approximation of the true solution. Our code then seeks  axisymmetric solutions of the full nonlinear system, \eqref{EQ:vertical_force_normalizedFULL} and \eqref{EQ:compatibility_normalizedFULL}, using this starting point as a guess for a relaxation method. Once a solution is found, the value of $\lambda_s$ is decreased slightly using the previous solution (at a larger $\lambda_s$) as the initial guess; this process continues automatically, so that the `continuation' is performed in the parameter $\lambda_s$. AUTO uses arc-length continuation, allowing it to identify the fold bifurcation that corresponds to snap-through. The inset of figure \ref{FIG:lambda_normalized}(a) shows a comparison between the bistability threshold computed from the finite element ABAQUS simulations, denoted by $\lambda_{d}^{T}(\alpha)$, and the corresponding result from the continuation analysis of the shallow shell theory, which we denote by $\lambda_{s}^T$ (independent of $\alpha$, since the shallow shell result corresponds to the limit $\alpha\to0$). For all values of the angle $\alpha\lesssim1.3$, we see that the critical value of $\lambda_s$ at which the shallow system becomes bistable is within $2\%$ of the threshold value of $\lambda_d$ determined from our ABAQUS simulations with deep shells.

Given the excellent agreement between the shallow shell prediction and ABAQUS simulations, we shall make use of the shallow shell theory in the remainder of this paper to explore properties of the transition from monostable to bistable. We begin by studying the role of Poisson's ratio in this problem, see figure \ref{FIG:lambda_normalized}(b) where the value of $\lambda_s$ at the transition, $\lambda_{s}^T$, is shown as a function of $\nu$. Our numerical results from the shallow-shell model show that the threshold value of $\lambda_s$ is an approximately linear function of $\nu$: $\lambda_{s}^T\approx1.44\nu+5.06$. The predictions of analytical works available in literature, specialized to the  case of spherical shells, are  compared with our numerical results in figure \ref{FIG:lambda_normalized}(b); we note that there is a qualitative disagreement with the analysis provided by Seffen \& Vidoli \cite{Seffen2016_sms} (see the introduction of ref.~\cite{Seffen2016_sms}), while the analysis reported by Seffen \cite{Seffen2007_prsa} (see the first relation in Equation (3.5) of ref.~\cite{Seffen2007_prsa}) shows an increasing trend very close to our finding. We also emphasize that these analytical predictions were predicated on the assumption that the everted shape has a constant curvature --- an assumption that is not made in our numerical simulations, and which is seen to be especially dubious close to the transition between monostability and bistability. These limitations have recently been overcome by Sobota and Seffen \cite{Sobota2017_prsa} who use a Rayleigh--Ritz method to estimate the transition in terms of the critical shell indentation at which snap-through occurs.

Numerical results for the transition  value of $\lambda_d$ were obtained  by Brinkmeyer {\it et al.} \cite{Brinkmeyer2012_ijss}; they focused on Poisson ratios $0.4\leq\nu\leq0.49$ and proposed that in this case the transition happens when the modified parameter $\lambda_{br} = 12^{1/4}\bigl(1-\nu^2\bigr)^{n/4} \alpha (R/H)^{1/2}\approx5.34$ with $n=1.85$ a fitted parameter. Figure \ref{FIG:lambda_normalized}(b) shows that this criterion is in reasonable agreement with our numerics in the region of $\nu$ for which the fitting was performed, but quickly deviates outside this region.

In summary, we find that for all shells up to hemispherical shells, the threshold between bistability and monostability is well approximated by a single critical value of $\lambda_d$ that depends only on the Poisson ratio $\nu$ --- the critical value $\lambda_d^T(\alpha)$ is approximately constant, and independent of $\alpha$. Furthermore, this critical value is very close to the critical value of the parameter $\lambda_s$: by setting $\lambda_d^T(\alpha)\approx\lambda_s^T$ we have an approximation that appropriately captures the dependence of the threshold on the shell depth, $\alpha$, with an error of less than $1\%$ for  $\alpha \lesssim 1.12\mathrm{~rad}$, and less than $2\%$ for $\alpha\lesssim1.3\mathrm{~rad}$. 

Finally, we note that our numerical results are consistent with the computations of Mescal \cite{Mescall1965_jam} that are recalled by Libai and Simmonds \cite{LibaiSimmonds} to infer an effective value of $\lambda^{(cr)}$. However, we can now provide a more definitive statement that, for a given Poisson's ratio, the numerically computed threshold for bistability is $\lambda_d^T\approx\lambda_s^T$  for shells as deep as $L/R \approx 0.97$, or $\alpha\approx 1.3$.

Having studied this transition for unforced shells, we now move on to study the robustness of this threshold. In particular, we seek to understand how and when the transition between two stable states can be induced by the application of a point-like force.

\section{Induced snap-through in bistable shells}\label{SEC:Indentation}
In the previous Section, we showed that stable everted configurations exist only for spherical caps with a large enough value of the parameter $\lambda\propto (R/h)^{1/2}\alpha$. However, there are two aspects regarding the robustness of this transition from monostability to bistability that are not clear: first, we expect that close to the transition, simply touching the everted shape should be enough to cause the shell to rapidly snap to its natural state. Second, while mirror buckling represents a natural deformation of thin shells (see, for example, \cite{Pogorelov,Gomez2016_prs}), it is also well known that in the pointwise indentation of a spherical thin shell one quickly finds that the shell develops a polygonal buckling pattern \cite{Vaziri2008_pnas, Vella2011_prl,Nasto2014_jam}.

To address these two aspects of the robustness of bistability, we now consider how the application of a point load (indentation) changes the picture outlined in the previous section, and address the questions of when bistability and axisymmetry are lost via the application of a point force. To answer these questions we shall only make use of experiments and shallow-shell theory --- our work in \S\ref{SEC:Bistability} has shown that the simplified shallow-shell approach has a surprisingly large domain of validity. In what follows we shall also drop the subscript `$s$' from $\lambda$, since we expect our results to be largely replicated for the value of $\lambda$ appropriate to deep shells.

Related questions were considered by Fitch \cite{Fitch1968_ijss} for shallow spherical caps with clamped edges. Fitch demonstrated that, depending on the parameter $\lambda$, such shells may undergo either an asymmetric polygonal buckling or reach a maximum force (so that increasing the indentation depth decreases the force). This latter scenario was referred to as `axisymmetric snapping' by Fitch \cite{Fitch1968_ijss}, but only corresponds to snap-through to an everted state if the experiment is conducted with a controlled load (i.e.~force-controlled). In our experiment, which corresponds to displacement control, \emph{snapping} occurs when the shell reaches a self-equilibrated, but unstable, equilibrium. Such a snap-through never occurs in the framework investigated by Fitch \cite{Fitch1968_ijss}, since the clamped boundary conditions used there do not allow the everted state to exist without an applied force. Nevertheless, the analysis of Fitch \cite{Fitch1968_ijss} provides a useful point of comparison, and is discussed in Appendix \ref{APP:NaturalShells}.

\subsection{Axisymmetric Indentation and Snap-Through}

We investigated indentation by incorporating a point force as a source term on the right-hand side of \eqref{EQ:force_balance}, $-F\delta(r)/(2\pi r)$. Here $\delta(\cdot)$ denotes the Dirac $\delta$-function, and corresponds to an ideal, `point-like' indenter. We also introduce the dimensionless indentation depth, $\wo$, and indentation force, $\cF$, defined by
\begin{equation}
\wo=\frac{\delta_0R}{L^2},\quad \cF=\frac{F R^3}{L^4 Eh},
\end{equation}
where $\delta_0$ is the dimensional displacement applied to the shell apex. The resulting axisymmetric equations can be solved numerically using, for example, the MATLAB routine \texttt{bvp4c}, and give the force $\cF(\wo)$ required to induce a given indentation depth $\wo$.

\begin{figure}
\centering
\begin{subfigure}{0.49 \linewidth}
	\includegraphics[width=0.95\linewidth]{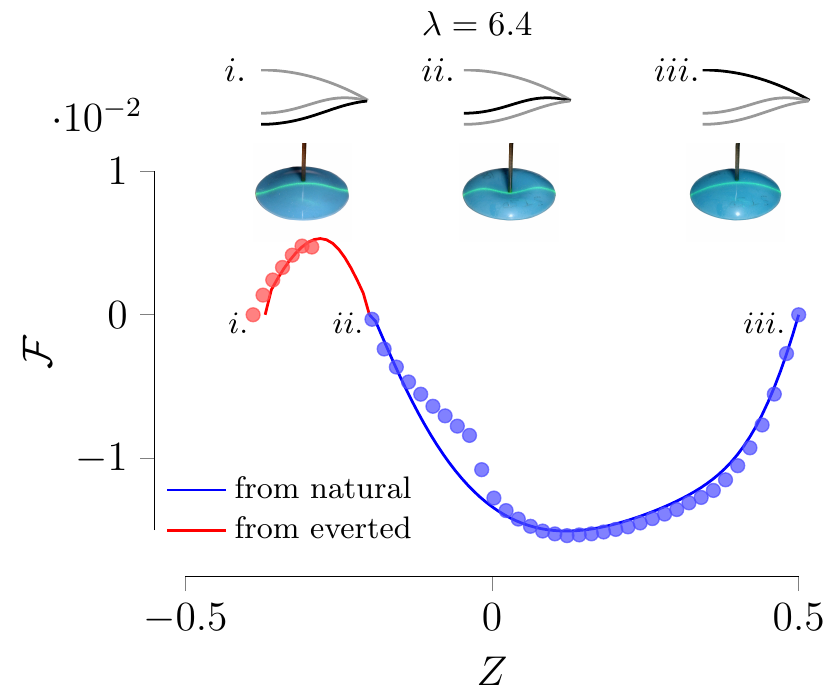}	
	\caption{}
\end{subfigure}
\begin{subfigure}{0.49 \linewidth}
	\includegraphics[width=0.95\linewidth]{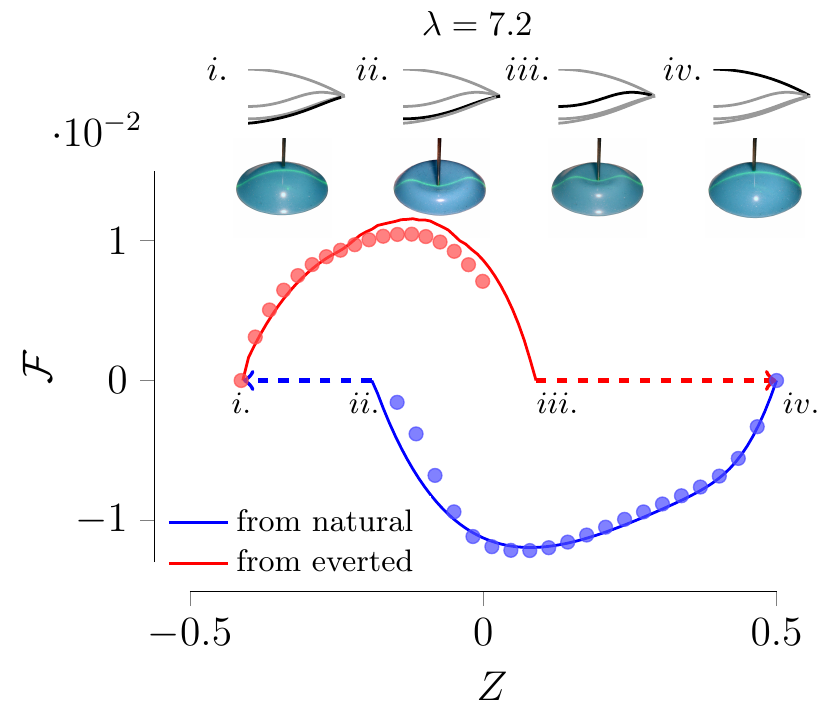}
	\caption{}
\end{subfigure}
\begin{subfigure}{0.49 \linewidth}
	\includegraphics[width=0.95\linewidth]{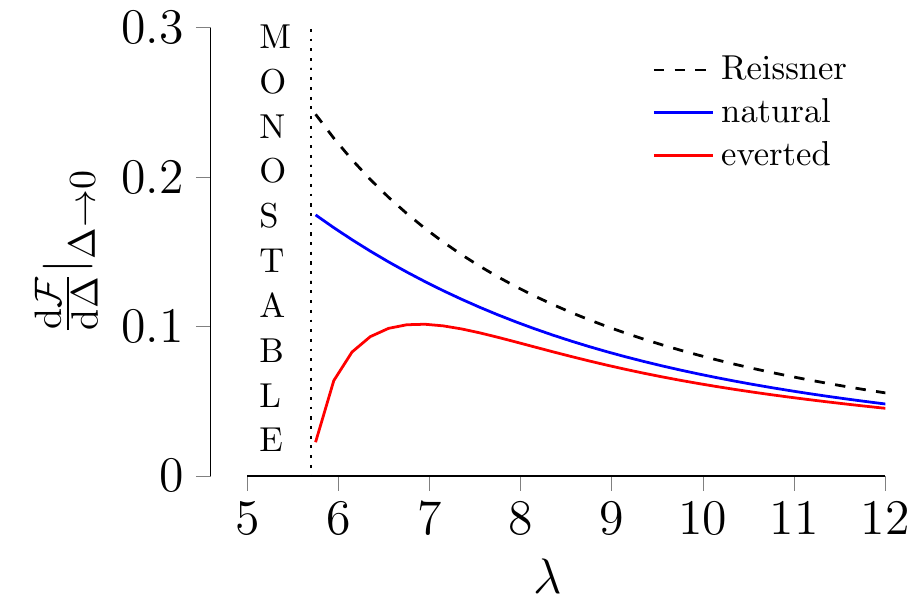}
	\caption{}
\end{subfigure}
\begin{subfigure}{0.49 \linewidth}
	\includegraphics[width=0.95\linewidth]{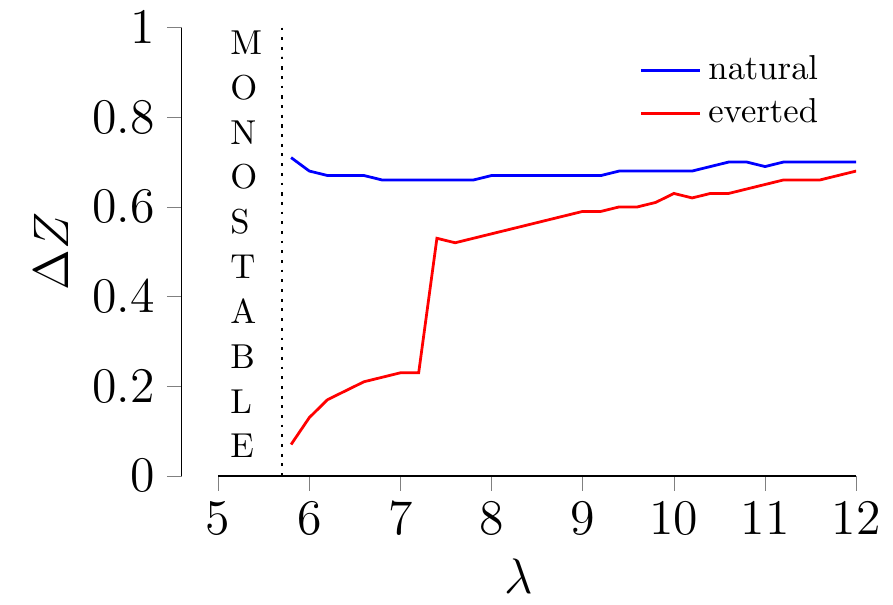}
	\caption{}
\end{subfigure}
\caption{The force--displacement relationship for the indentation of a shell from its natural (red curves and circles) and everted (blue curves and circles) shapes. Numerically determined data are represented by continuous curves, with experimental data represented by points. Data are shown for shells close to the transition between bistability and monostability (which occurs at $\lambda\approx 5.75$ for $\nu=0.5$): (a) $\lambda = 6.4$ while in (b) $\lambda=7.2$. Deformed shapes with $\mathcal{F}=0$ are shown in the insets: in each case the highlighted shape is that expected at the corresponding Roman numeral on the graph. (c) Tangent stiffness at zero displacement ($\Delta \rightarrow 0$) in the natural configuration (blue) and the everted  configuration (red).  For sake of comparison the two identified domains of axisymmetric snapping and hysteretic snapping are highlighted. The classic result for an unpressurized shell ${\cal F}=8 \Delta /\lambda^2$ due to Reissner \cite{reissner47a,reissner47b} is shown as the dashed curve. (d) The \emph{robustness} parameter $\Delta Z$, the dimensionless displacement at the point of snapping (measured with respect to the natural or the everted self-equilibrated state, as appropriate), shows that the displacement required to force a shell to snap-through from its natural configuration remains fairly independent of the shell thickness (blue curve), while the robustness of the everted shell to snap-through is highly dependent on the shell thickness, through the parameter $\lambda$ (red curve).} \label{FIG:SYMvsASYM_snapping}
\end{figure}

Figure \ref{FIG:SYMvsASYM_snapping} shows the numerically computed force--indentation curves for two different values of $\lambda$ close to the unforced stability transition; this numerical analysis is similar to that of Broadland \& Cohen \cite{Brodland1987_ijss}, though they did not focus on the robustness of the unforced transition. For each value of $\lambda$, the force $\mathcal{F}$ shows a non-monotonic behaviour as the shell is poked away from its natural  or everted  state (and towards the other state): the absolute value of the indentation force initially increases as the apex position, $Z$, changes from its initial value (which we denote by $Z_N$ and $Z_E$ for the natural and everted states, respectively). However, the force $ \mathcal{F}$ reaches a maximum and then decreases to zero, at which point snap-through occurs at apex position $Z_S$. This is qualitatively similar to the behaviour in the snap-through of arches investigated by Pandey {\it et al.} \cite{Pandey2014_epl}. However, we note that the value of $\lambda$ does qualitatively change the behaviour of the system: with $\lambda=6.4$  we find that pushing from either of the everted or natural shapes leads to force displacement curves that meet at a zero force solution, with a continuous axisymmetric path joining the two (in practice snap-through between the states occurs at this intermediate configuration with zero force, since the indenter is not adhesive). We refer to this regime as the `axisymmetric snapping' regime. In contrast, with $\lambda=7.2$ different axisymmetric solutions exist (depending on whether one starts from the everted or natural shape). Snap-through again occurs at the point of zero force, $\cF=0$, and we refer to this as `hysteretic snapping' since the indentation depth at which snap-through occurs is history dependent. (We will discuss in detail these two regimes after we introduce a third regime, `axisymmetric buckling'.) We also note that the amount of indentation from the everted shape required to induce snapping increases as $\lambda$ increases away from the bistability transition: the shell becomes more robust to snap-through, as might be expected. This indentation robustness $\Delta Z$  (defined by $\Delta Z = |Z_N-Z_S|$ and $\Delta Z = |Z_E-Z_S|$, for indentation from natural and everted states, respectively) is plotted as a function of $\lambda$ in figure \ref{FIG:SYMvsASYM_snapping}d.

Figure \ref{FIG:SYMvsASYM_snapping} also shows experimental data (points) for comparison with the numerical results (curves). While the quantitative agreement between the two is generally good, we note that there is an important distinction  for small $\lambda$: experimentally, contact is lost very soon after the maximum value of $\mathcal{F}$ is reached (see the red branch in figure \ref{FIG:SYMvsASYM_snapping}(a)). We attribute this premature snap-through to a very steep decrease in the force--displacement curve that is not captured with the simple indentation model. We recall that we have assumed a simply supported edge in determining the numerical results presented in figure \ref{FIG:SYMvsASYM_snapping}. If one accounts for the fact that the outer contact point between the shell and the substrate may change with indentation depth, this discrepancy is largely resolved (see Appendix \ref{APP:wall}). 

The force--displacement relationships in figure \ref{FIG:SYMvsASYM_snapping} show two further features of the indentation of natural and everted shells that are pertinent to our discussion of the robustness of bistability: (i) the imposed indentation  required to cause an everted shell to snap increases with $\lambda$ and (ii) an everted shell is `softer' than the corresponding natural shell with the same value of $\lambda$. Both of these observations are intuitive (shells with smaller $\lambda$ are closer to the transition and hence `easier' to snap), while everted shells are `trying' to return to their unstressed state and hence take less force to deform them in this direction. Figure \ref{FIG:SYMvsASYM_snapping}c and figure \ref{FIG:SYMvsASYM_snapping}d quantify both of these observations further.

\subsection{Buckling Analysis}

The preceding analysis showed that the transition between everted and natural shapes may be controlled by the application of a localized force. It is then natural to investigate whether the axisymmetric and hysteretic snapping discussed in the previous section are observed in reality: does the shell buckle asymmetrically prior to snap-through?

To answer this question, we perturb the axisymmetric base state with an azimuthal ($\theta$) dependence that is motivated by the observation of buckling in this direction \cite{Fitch1968_ijss,Pauchard1998_PMb,Vaziri2008_pnas,Vella2011_prl}. We now denote the axisymmetric quantities by superscripts $(0)$ and the corresponding perturbed quantities at first order by superscripts $(1)$, and make an {\em ansatz} of the form:
\begin{equation}\begin{split}\label{EQ:linearperturbation}
W\left(\rho; \theta\right)= W^{(0)}\left(\rho\right) + W^{(1)}\left(\rho\right)\cos\left(m \theta\right)\\
\Phi\left(\rho; \theta\right)= \Phi^{(0)}\left(\rho\right) + \Phi^{(1)}\left(\rho\right)\cos\left(m \theta\right). 
\end{split}
\end{equation}

Details of the buckling analysis are given in Appendix \ref{APP:LinearStability}. To allow a comparison with the earlier work of Fitch \cite{Fitch1968_ijss}, Appendix \ref{APP:NaturalShells} considers the effect of clamped boundary conditions on the buckling of such shells. However, for these boundary conditions, shells are not bistable --- since our interest is in the robustness of bistability, in the main text we therefore focus on characterizing the  onset of asymmetrical buckling for different values of the parameter $\lambda\gtrsim5.75$, for which we have already seen that free, unloaded shells are bistable. Further, we restrict attention to the case of incompressible shells, $\nu=0.5$.

\begin{figure}[h]
\centering
\begin{subfigure}{0.49 \linewidth}
	\includegraphics[width=0.95\linewidth]{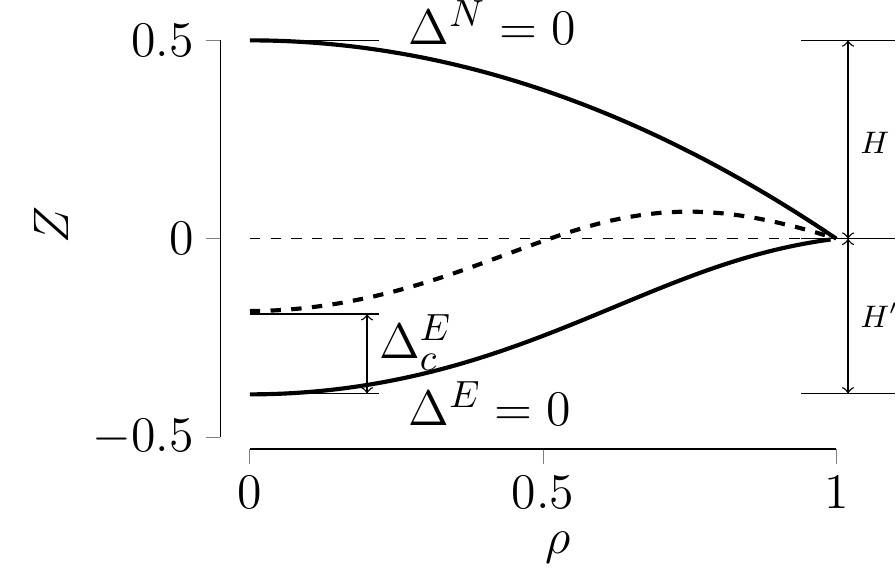}	
	\caption{}
\end{subfigure}
\begin{subfigure}{0.49 \linewidth}
	\includegraphics[width=0.95\linewidth]{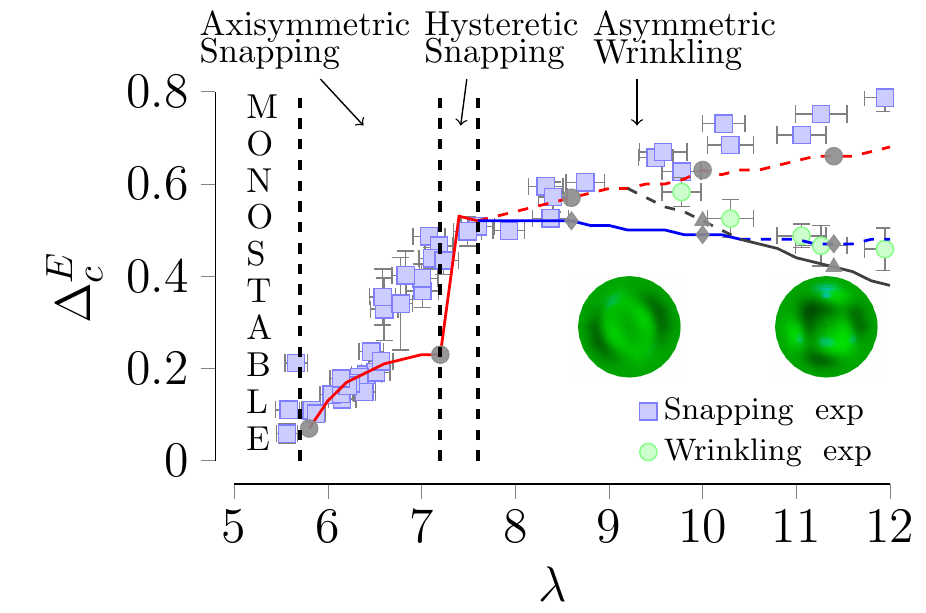}
	\caption{}
\end{subfigure}
\caption{(a) Sketch of the equilibrium shapes in the axisymmetric snapping of a spherical shell. The two continuous curves represent the natural and the everted stable configurations, while the dashed curve is the unstable equilibrium configurations with zero applied force. Accordingly, $H$ and $H'$ are the rise of the natural and everted shell with respect to the position of the outer edge. (b) Indentation of an everted shell according to the shallow shell theory. Snapping is identified by the red lines with grey circles while the  thresholds for asymmetric buckling with $m=2$ and $m=3$ are represented by blue (with diamonds) and grey (with triangles) lines, respectively. For each condition, the line is continuous if that behaviour emerges (i.e.~has the smallest indentation $\wo_c^E$ of all events), and is dashed otherwise. Experimental results are also reported: snapping or wrinkling are shown as blue and square markers and green circles, respectively. Note, in particular, that even in the asymmetric wrinkling regime, the experimentally observed snap-through transition is well approximated by the axisymmetric theoretical prediction for snap-through (red dashed curve). } \label{FIG:PhaseDiagram_IndentationEverted}
\end{figure}

Figure \ref{FIG:PhaseDiagram_IndentationEverted} presents numerical results for the indentation of an everted spherical cap, together with experimental results in the same condition. We note that asymmetric buckling is the preferred deformation for thinner shells (large $\lambda$), as we will see also for indentation of the natural case. Furthermore, such shells also show an increasing mode number at onset of instability as $\lambda$ increases (corresponding to thinner shells), saturating at a mode number $m=3$ as $\lambda\to\infty$. (This ties in with the case of a free, natural shell for which it is known\cite{Pauchard1998_PMb,Vaziri2008_pnas,Vella2011_prl} that buckling is observed first with $m=3$, while it runs counter to the everted, clamped case \cite{Fitch1968_ijss}, as shown in Appendix \ref{APP:NaturalShells}.) One surprising feature of the experimental results is that the purely axisymmetric theory is able to predict the final snap-through indentation even in cases where the shell has buckled azimuthally so that the axisymmetric theory is no longer strictly valid (the dashed red curve approximately passes through the blue squares in the asymmetric wrinkling region of figure \ref{FIG:PhaseDiagram_IndentationEverted}).

A notable feature of the results shown in figure \ref{PhaseDiagram_IndentationEverted}(b) is the critical indentation depth at which snap-through is observed, $\Delta_c^E$, is generally a smooth function of $\lambda$ but suddenly increases in the range $7.2\lesssim\lambda\lesssim7.6$. This jump is evident in both our numerical results and experiments, though experiments show a smoother transition than our theoretical calculations;  this difference is reduced  if the details of contact with the substrate are included (see Appendix \ref{APP:wall}). We attribute the transition in the values of $\Delta_c^E$  to a transition between two deformation mechanisms: for large $\lambda$ (relatively thin shells) global deformation occurs via bending rather than stretching, while for small $\lambda$ (relatively thick shells) deformation occurs via a combination of bending \emph{and} stretching.

\subsection{A regime diagram}

\begin{figure}
\centering
\includegraphics[width=0.66\columnwidth]{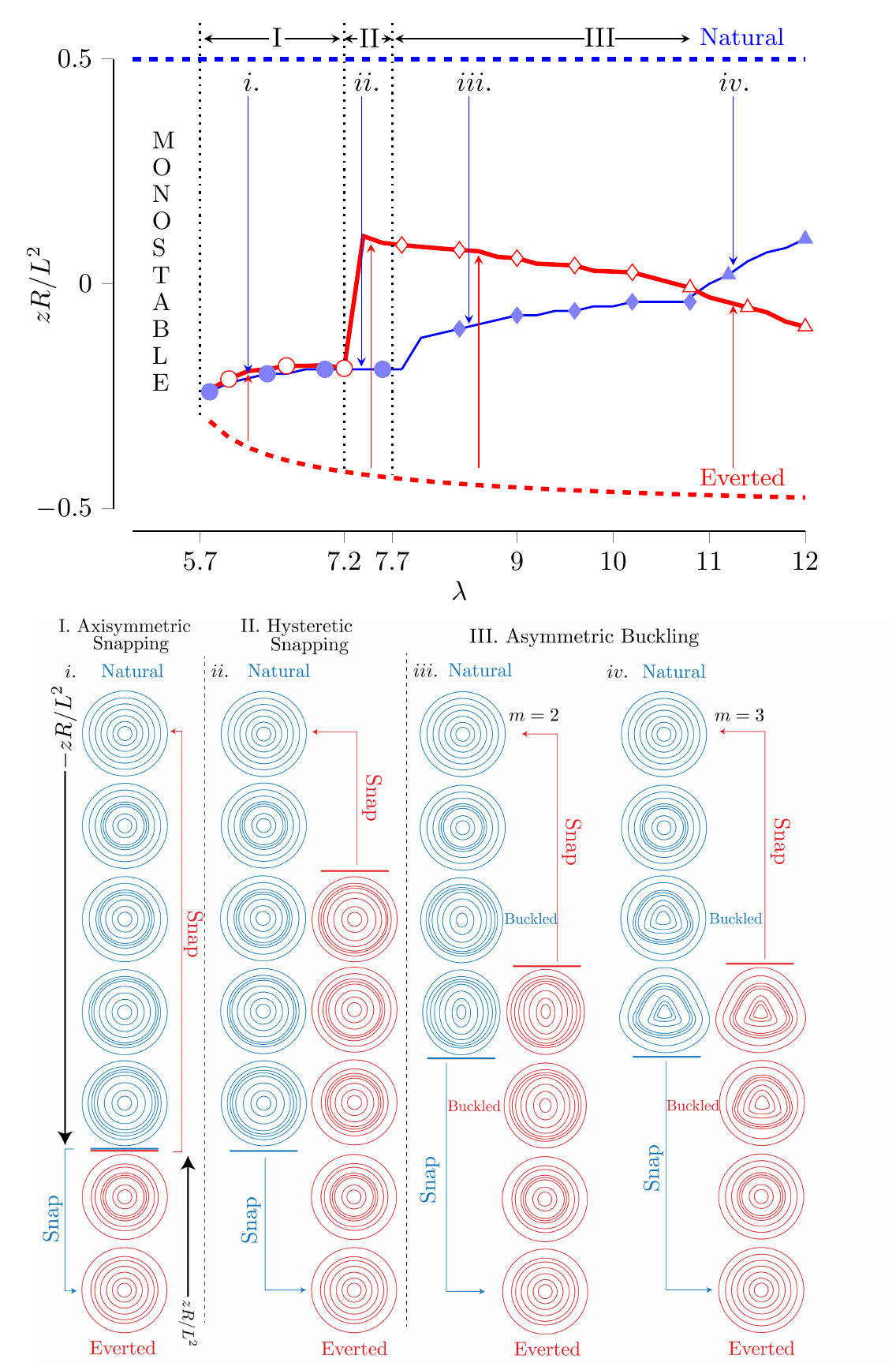}
\caption{Top panel: Regime diagram showing the behaviour of indented shells ($\nu=0.5$) with free edges. The thick red curve with open markers identifies the critical (either snapping or buckling) apex position  for the indentation of an everted shell; the thin blue curve with filled markers refers to indentation from the natural state. Different behaviours are indicated by different symbols: snapping (circles), buckling with $m=2$ (diamonds) and buckling with $m=3$ (triangles). The position of the apex in the self-equilibrated states is indicated for the natural configuration ($Z = z R/L^2=0.5$, blue dashed line) and for the everted configuration (red dashed line). Thin vertical arrows indicate the direction of the applied indentation in the two cases. Lower panel: Qualitative representations of the shapes of the shell (top view) for the three different regimes identified in the bistable domain. In the axisymmetric snapping case, snap-through happens with the same apex position (regardless of the initial state) so that there is a continuous path between natural and everted states; in the hysteretic snapping case, shells remain axisymmetric until the snap-through point but the apex position is different for the two initial conditions; in the asymmetric buckling case, the shell  buckles, with $m=2$ or $m=3$ before snapping occurs, for both natural and everted shells.}
\label{FIG:phase_diagram_theory}       
\end{figure}

Considering also the behaviour of an indented free, natural shell we may summarize our numerical results in a regime diagram, see figure \ref{FIG:phase_diagram_theory}. Here the key parameters are the position of the apex of the shell relative to its edges and the shell depth parameter, $\lambda$. We note that there are four qualitatively different behaviours are observed, depending on the value of $\lambda$:

\begin{itemize}

\item $\lambda \lesssim 5.75$: MONOSTABLE. Here, the only zero-force equilibrium of the shell is the natural state; as soon as the indentation force is removed, the shell immediately returns to its natural state. 

\item $5.75\lesssim\lambda \lesssim 7.2$: AXISYMMETRIC SNAPPING. Both everted and natural states exist. Furthermore, one may transition between these two states via snap-through at an intermediate (unstable) self-equilibrated state ($\cF=0$). The indentation required to effect this transition is the same (to within numerical errors) whether one starts with an everted or natural state.

\item $7.2\lesssim\lambda\lesssim7.7$: HYSTERETIC SNAPPING. Both everted and natural states exist. The transition between them occurs via snap-through at an intermediate (unstable) self-equilibrated state ($\cF=0$). However, in contrast with axisymmetric snapping, the indentation depth at which the snap-through transition occurs is path-dependent, depending on whether one starts with an everted or natural shell \cite{Brodland1987_ijss}. 

\item $\lambda \gtrsim 7.7$: ASYMMETRIC BUCKLING. For sufficiently thin shells (sufficiently large $\lambda$), both the natural and the everted shells exhibit azimuthal buckling before  snap-through occurs. Note also that the axisymmetric theory appears to give a good account of the experimentally observed transition to snap-through, even though this theory it is not strictly valid after buckling.
\end{itemize}

\section{Conclusions\label{SEC:Conclusion}}
In this paper, we have considered different aspects of the  bistability of a spherical shell cap. We began by determining the transition from bistability to monostability and showed that an approximate condition for a spherical shallow shell to be  bistable is that
\beq
\left[12 \left(1-\nu^2\right)\right]^{1/4}\frac{L}{(hR)^{1/2}}=\lambda_s\geq \lambda_s^T(\nu)\approx 1.44\nu+5.06
\label{eqn:BistabThreshold}
\eeq with $\nu$ the  Poisson's ratio. This criterion was determined from numerical solutions of the shallow shell equations and so is strictly only valid for $L/R\to0$. However, comparison with finite element simulations for specific values of $\nu$, and a wide range of shell angles $\alpha=\sin^{-1}(L/R)$, as well as experiments with $\nu=0.5$, showed that this condition is unreasonably effective: relatively deep shells are bistable provided that $\lambda_d\gtrsim\lambda_s^T(\nu)$ and  $L/R\lesssim0.97$, or $\alpha\lesssim 1.3$.   Furthermore, our numerical results provided a more extensive coverage of the effect of Poisson's ratio than was possible in previous numerical simulations \cite{Brinkmeyer2012_ijss}. We emphasize that the dependence of the threshold for bistability on only two dimensionless parameters, $\lambda_s$ and $\nu$, in \eqref{eqn:BistabThreshold} is related to the approximations involved in the DMV theory, which is valid only for shallow shells with moderate rotations. For deeper shells, which we have established means $\alpha\gtrsim1.3$, the critical value of $\lambda_d$ at which bistability is lost will depend on $\alpha$ as well as $\nu$ (as shown in figure \ref{FIG:lambda_normalized}). Equivalently, there is a dependence on the dimensionless shell thickness $h/R$ since $\alpha$, $\nu$ and $h/R$ all enter the definition of $\lambda_d$ in \eqref{EQ:lambda_d}. We have not investigated the behaviour of the critical value of $\lambda_d$ for $\alpha\gtrsim 1.3$ here, but this would be a worthwhile direction for future research.

Having seen that shallow shell theory provides a good account of the transition between self-equilibrated bistability and monostability in the absence of an externally applied load, we then moved on to investigate the robustness of this transition. To do this we used a combination of shallow shell theory and experiments. The results of our analysis is the regime diagram in figure \ref{FIG:phase_diagram_theory}, which shows the different regimes of the $(\lambda,Z)$ parameter space for which the different behaviours are observed. In particular, we observe that for intermediate shell depths, $5.75\lesssim\lambda\lesssim 7.7$, the (otherwise bistable) shell can be induced to snap-through by indentation axisymmetrically (i.e.~without any azimuthal buckling). For still deeper shells, the shell buckles before snapping; the wavenumber observed at buckling depends on the shell depth (as well as the boundary conditions). Qualitatively the same behaviour is observed for everted shells as for natural shells. 

Our results show that the everted state becomes more robust to eversion as the shell becomes deeper, as might be expected, but ultimately that this everted state gives rise to an azimuthal buckling instability. Surprisingly, the purely axisymmetric theory appears to give a good account of \emph{when} snap-through occurs, even beyond this buckling instability. We believe that this robustness of the axisymmetric snap-through transition to azimuthal buckling suggests that the underlying mirror buckled solution, while not ultimately attained in this limit, heavily influences the stability of the whole shell structure.

\vskip6pt

\section*{Acknowledgment}

The research leading to these results has received funding from the European Research Council under the European Union's Horizon 2020 Programme/ERC grant agreement no. 637334 (D.V.). D.P.H.~and X.J.~are grateful for financial support from the NSF CMMI--1505125. We thank  Mariano Galvagno for  discussions and preliminary numerical investigations. Data associated with this paper may be found at \texttt{https://doi.org/10.5287/bodleian:O5w4Berw0}.

\vskip2pc

\clearpage
\appendix       
\section{Materials and Methods}\label{APP:MatMet}
\subsection{Experiments}
To investigate which geometric properties govern the bistability of deep shells, we fabricated shells with a variety of  dimensions.  In our experiments, spherical caps were cast from polyvinylsiloxane rubber (PVS, Zhermack Elite Double 32F) using the coating method \cite{brun2016}. The radii of curvature of these spherical shells was varied in the interval $13\mathrm{~mm} \lesssim R\lesssim 38\mathrm{mm}$ by coating on steel spheres of these radii. The viscosity of the mixed material increases through time so that the thickness of the spherical shell can be controlled by changing the waiting time before pouring the liquid material onto the spherical ball \cite{Knoche2014_epj}. To get even thicker shells, we can repeat the process and apply additional layers on top of the existing spherical caps. After the polymer has cured, these shells were carefully cut and removed from the sphere on which they were cast. While the coating method leads to a shell of approximately uniform thickness for sufficiently small values of the  meridional angle $\alpha$, the thickness becomes highly non-uniform for $\alpha\gtrsim 1.3$; we therefore focus on shells with $\alpha\lesssim 1.3$ here.

To quantify the stability of a given spherical cap shell, point indentation tests were conducted using an INSTRON U5943 machine with a $5\mathrm{~N}$ load cell, which has a precision of $0.002$N. The loading rate used in the tests were $10\mathrm{~mm/min}$; an indenter with a fixed diameter of $1.6\mathrm{~mm}$ was used for all experiments. The everted spherical shell is placed on an acrylic plate with a hole cut in the center. A circle that is the same size as the boundary of the shell is drawn on the acrylic plate. We then calibrate the indenter to the center of the circle and place the everted spherical cap in the circle to ensure concentric loading. To minimize the effect of boundary friction between the outer edge of the shell and acrylic plate used as a substrate, we deposited a thin layer of silicone oil ($5\mathrm{~cSt}$, Sigma Aldrich). Each data point is the average of five tests.

\subsection{ABAQUS simulations}
We use finite element simulations to investigate the behaviour of deep shells: the model geometry shown in figure \ref{FIG:lambda_experiments}(a) is implemented in ABAQUS 6.14 (Dassault Syst{\'e}mes Simulia Corp., Providence, RI, USA). Since our primary interest lies in the axisymmetric bistability of spherical shells, the full three-dimensional geometry is reduced to a two-dimensional slice. This slice of the shell is discretized using axisymmetric continuum CAX4H elements and a hyperelastic, Neo-Hookean material is used. Following a sensitivity analysis, all simulations reported here were carried out with no fewer than 5000 hybrid elements. 
The simulations are conducted in two steps: first an indenter is used to apply a prescribed indentation depth, with contact between the shell and indenter enforced. The prescribed indentation depth is twice the initial rise of the cap. The indenter is then removed in the second step, allowing the system to find the closest equilibrium solution (all simulations are static, without any stabilization enforced).
Rigid motions are avoided by imposing vanishing displacement $u_{\mathrm{axi}}=0$ in the radial ($r$-direction) at the central axis of the shell (i.e.~for all elements that have $r=0$); similarly the vertical ($z$) displacement is required to vanish, $V_B=0$, only on the lower node of the mesh at the edge (i.e. $r=L$ and $z=0$). These conditions are consistent with simple support conditions and are represented graphically in figure \ref{FIG:lambda_experiments}(a). 

\section{Linear Stability Analysis}\label{APP:LinearStability}

Inserting the buckling \emph{ansatz} \eqref{EQ:linearperturbation} into the dimensionless shallow shell equations \eqref{EQ:vertical_force_normalizedFULL} and \eqref{EQ:compatibility_normalizedFULL} we find, after linearization, that the radial dependence of the perturbation (i.e.~the functions $\Phi^{(1)}$ and $W^{(1)}$)  satisfy:
\begin{equation}
\frac{1}{\lambda_s^4}\nabla_m^2\nabla_m^2 W^{(1)} + \nabla_m^2 \Phi^{(1)}- \Lop_m\left(\Phi^{(0)},W^{(1)}\right) - \Lop_m\left(W^{(0)},\Phi^{(1)}\right) = 0
\label{eqn:linpert1}
\end{equation}and 
\begin{equation}
\nabla_m^2\nabla_m^2 \Phi^{(1)} + \Lop_m\left(W^{(0)},W^{(1)}\right) - \nabla_m^2W^{(1)} = 0
\label{eqn:linpert2}
\end{equation} where
\begin{equation}
\nabla^2_m f=\frac{1}{\rho}\frac{\upd}{\upd\rho}\left(\rho \frac{\upd f}{\upd \rho}\right)-\frac{m^2}{\rho^2}f
\end{equation} and
\begin{equation}
\Lop_m(f,g)=\frac{\upd^2 f}{\upd \rho^2}\left(\frac{1}{\rho}\frac{\upd g}{\upd \rho} - \frac{m^2}{\rho^2} g\right)+\frac{\upd^2 g}{\upd \rho^2}\left(\frac{1}{\rho}\frac{\upd f}{\upd \rho} \right).
\end{equation}

We now seek to find the smallest indentation depth (for integer $m$) for which a non-trivial solution of equations \eqref{eqn:linpert1}--\eqref{eqn:linpert2} exists. This critical value depends on the axisymmetric base state $\left[W^{(0)}(\rho),\Phi^{(0)}(\rho)\right]$, which will differ between natural and everted configurations, as well as with the boundary conditions (\emph{i.e.}~clamped or free). The appropriate boundary conditions, together with their physical meaning, are summarized in table \ref{TAB:BCs_zeroth} (for the axisymmetric base state) and table \ref{TAB:BCs_linear} (for the linear perturbation problem). 

The critical indentation depth, $\Delta_c$, together with the wave number at the onset of instability, $m_c$, are determined numerically as follows. For perturbations from the natural configuration, the axisymmetric problem for a given $\wo$ is directly solved numerically in MATLAB using  the built-in function \texttt{bvp4c}. For perturbations from the everted configuration, we first determine the everted shape prior to indentation ($\wo=0$), before applying  a displacement $\wo$ to the apex of the everted shape.  In either case,  the functions $W^{(0)}\left(\rho;\wo\right)$ and $\Phi^{(0)}\left(\rho;\wo\right)$ are determined for a given value of $\wo$ in this manner. With  the axisymmetric base state computed, we proceed to the perturbed problem: the domain $\left[0,1\right]$ is sub-divided into $N$ grid points and equations \eqref{eqn:linpert1}--\eqref{eqn:linpert2} is discretized using a  second-order finite difference scheme (central differences); the appropriate  boundary conditions are also discretized using second-order accurate finite difference stencils.

The  discretization described above can then be rewritten as a quadratic eigenvalue problem
\begin{equation}
\left[{\bf A_0} + n {\bf A_1} + n^2 {\bf A_2}\right]\bf{X} =\bf{0}
\label{eqn:MatEqn}
\end{equation}
where $n=m^2$, the matrices ${\bf A_0}$, ${\bf A_1}$ and ${\bf A_2}$ depend only on the zeroth order problem, and the vector of unknowns ${\bf X} = \left[W^{(1)}_1 W^{(1)}_2 ... W^{(1)}_p ... W^{(1)}_N \Phi^{(1)}_1 \Phi^{(1)}_2 ... \Phi^{(1)}_q ... \Phi^{(1)}_N \right]^T$ contains the $N$ nodal values of the perturbation to the displacement and the Airy stress function. 

\begin{table}[t]
	\caption{The six boundary conditions required to solve the axisimmetric base state ({\em i.e.}~for $\Phi^{(0)}$ and $W^{(0)}$). Different boundary conditions are appropriate depending on whether the boundary is free (denoted `F' below) or clamped (denoted `C' below). At the zeroth order, the displacement in the normal direction is denoted by $W^{(0)}$ while $U_r^{(0)}$ is the in-plane radial displacement. In-plane radial stress is indicated by $\Sigma_{rr}^{(0)}$ and $M_r^{(0)}$ is the bending moment around the hoop direction according to the notation of Ventsel and Krauthammer \cite{Ventsel2001}.}
	\begin{center}
		\label{TAB:BCs_zeroth}
		\begin{tabular}{p{0.06\linewidth} p{0.28\linewidth} p{0.43\linewidth} p{0.13\linewidth}}
			& & \\ 
			\hline
			\multicolumn{4}{|c|}{Boundary Conditions for the zeroth order problem} \\
			\hline
			\multirow{1}{*}{$\rho=0$} & Imposed normal displacement & $W^{(0)}(0)=\wo$ & F/C \\ 
			\multirow{1}{*}{$\rho=0$} & Zero radial displacement & $U_r^{(0)}(0)=\lim\limits_{\rho \rightarrow 0}\left[\rho \frac{\upd^2 \Phi^{(0)}}{\upd \rho^2} - \nu \frac{\upd \Phi^{(0)}}{\upd \rho} \right] =0$ & F/C\\
			\multirow{1}{*}{$\rho=0$} & Avoid cusp & $\frac{\upd W^{(0)}}{\upd \rho}\bigg\rvert_{\rho=0}=0$ & F/C\\
			\multirow{1}{*}{$\rho=1$} & $\begin{cases}
			\mathrm{Zero \; radial \; stress}\\
			\mathrm{Zero \; radial \; displacement}
			\end{cases}$
			
			& $\begin{cases} \Srr^{(0)}(0) = \left[\frac{1}{\rho}\frac{\upd \Phi^{(0)}}{\upd \rho}\right]\bigg\rvert_{\rho=1}=0\\
			U_r^{(0)}(1)= \left[\rho \frac{\upd^2 \Phi^{(0)}}{\upd \rho^2} - \nu \frac{\upd \Phi^{(0)}}{\upd \rho}\right]\bigg\rvert_{\rho=1} =0
			\end{cases}$ & $\begin{cases}
			\mathrm{F}\\
			\mathrm{C}
			\end{cases}$ \\
			\multirow{1}{*}{$\rho=1$} & $\begin{cases}
			\mathrm{Zero \; bending \; moment}\\
			\mathrm{Zero \; slope}
			\end{cases}$
			& $\begin{cases} M_r^{(0)}(1)=\left[\frac{\upd^2 W^{(0)}}{\upd\rho^2} + \frac{\nu}{\rho} \frac{\upd W^{(0)}}{\upd\rho}\right]\bigg\rvert_{\rho=1}=0\\
			\frac{\upd W^{(0)}}{\upd \rho}\bigg\rvert_{\rho=1}=0
			\end{cases}$ & $\begin{cases}
			\mathrm{F}\\
			\mathrm{C}
			\end{cases}$ \\
			\multirow{1}{*}{$\rho=1$} & Zero normal displacement & $W^{(0)}(1)=0$ & F/C \\
			\hline
		\end{tabular}
	\end{center}
\end{table}

\begin{table}[t]
	\caption{The eight boundary conditions required to solve the linear problem, presented for both the case of free boundary (F) and clamped boundary (C). At this linear order, the displacement in the normal direction is denoted by $W^{(1)}$ while $U_r^{(1)}$ and $U_{\theta}^{(1)}$ are the in-plane radial and hoop displacements, respectively. In-plane stresses are indicated by $\Sigma_{rr}^{(1)}$, $\Sigma_{r \theta}^{(1)}$. $M_r^{(1)}$ indicates the bending moment around the hoop direction according to the notation of Ventsel and Krauthammer \cite{Ventsel2001}.}
	\begin{center}
		\label{TAB:BCs_linear}
		\begin{tabular}{p{0.04\linewidth} p{0.22\linewidth} p{0.60\linewidth} p{0.05\linewidth}}
			& & \\ 
			\hline
			\multicolumn{4}{|c|}{Boundary Conditions for the linear problem} \\
			\hline
			\multirow{1}{*}{$\rho=0$} & Zero normal displacement & $W^{(1)}(0)=0$ & F/C\\
			\multirow{1}{*}{$\rho=0$} & Zero radial displacement & $U_r^{(1)}(0)=\lim\limits_{\rho \rightarrow 0}\left[\rho \frac{\upd^2 \Phi^{(1)}}{\upd \rho^2} - \nu \frac{\upd \Phi^{(1)}}{\upd \rho} + m^2\frac{\nu}{\rho} \Phi^{(1)} \right] =0$ & F/C \\
			\multirow{1}{*}{$\rho=0$} & Avoid cusp & $\frac{\upd W^{(1)}}{\upd \rho}\bigg\rvert_{\rho=0}=0$ & F/C \\
			\multirow{1}{*}{$\rho=0$} & Zero radial displacement ($ \propto m^2$)& $\Phi^{(1)}(0) =0$  & F/C \\
			\multirow{1}{*}{$\rho=1$} & $\begin{cases}
			\mathrm{Zero \; bending \; moment}\\
			\mathrm{Zero \; slope}
			\end{cases}$
			& $\begin{cases} M_r^{(1)}(1)=\left[\frac{\upd^2 W^{(1)}}{\upd\rho^2} + \nu\left(\frac{1}{\rho} \frac{\upd W^{(1)}}{\upd\rho}  -\frac{m^2}{\rho^2} W^{(1)}\right)\right]\bigg\rvert_{\rho=1}=0\\
			\frac{\upd W^{(1)}}{\upd \rho}\bigg\rvert_{\rho=1}=0
			\end{cases}$ & $\begin{cases}
			\mathrm{F}\\
			\mathrm{C}
			\end{cases}$ \\
			\multirow{1}{*}{$\rho=1$} & $\begin{cases}
			\mathrm{Zero \; radial \; stress}\\
			\mathrm{Zero \; radial \; disp}
			\end{cases}$
			& $\begin{cases}\Srr^{(1)}(1)=\left[\frac{1}{\rho}\frac{\upd \Phi^{(1)}}{\upd \rho} - \frac{m^2}{\rho^2}\Phi^{(1)}\right]\bigg\rvert_{\rho=1}=0\\
			U_r^{(1)}(1)= \left[\rho \frac{\upd^2 \Phi^{(1)}}{\upd \rho^2} - \nu \frac{\upd \Phi^{(1)}}{\upd \rho} + m^2\frac{\nu}{\rho}\Phi^{(1)}\right]\bigg\rvert_{\rho=1} =0\end{cases}$ & $\begin{cases}
			\mathrm{F}\\
			\mathrm{C}
			\end{cases}$ \\
			\multirow{1}{*}{$\rho=1$} & $\begin{cases}
			\mathrm{Zero \; shear \; stress}\\
			\mathrm{Zero \; azimuthal \; disp}
			\end{cases}$
			& $\begin{cases}\Srq^{(1)}(1)=\left[-\frac{\upd}{\upd \rho} \left(-\frac{m}{\rho}\Phi^{(1)}\right)\right]\bigg\rvert_{\rho=1}=0\\
			U_{\theta}^{(1)}(1)= \left[\rho\frac{\upd^3 \Phi^{(1)}}{\upd \rho^3} -\frac{1}{\rho} \left(1-\nu+m^2\left(2+\nu \right)\right) - \right. \\
			\quad\quad\quad\quad\quad\left.3\frac{m^2}{\rho^2}\Phi^{(1)} \right]\bigg\rvert_{\rho=1} =0\end{cases}$ & $\begin{cases}
			\mathrm{F}\\
			\mathrm{C}
			\end{cases}$ \\
			\multirow{1}{*}{$\rho=1$} & Zero Normal displacement & $W^{(1)}\left(1\right)=0$  & F/C \\
			\hline
		\end{tabular}
	\end{center}
\end{table}

We are only concerned with whether a non-trivial solution of the matrix equation \eqref{eqn:MatEqn} exists for a given integer $m$ and at what indentation depth $\wo$ such a solution exists ($\wo$ enters the problem through the base state functions $W^{(0)}(\rho;\wo)$ and $\Phi^{(0)}(\rho;\wo)$). For a given $\bar{m} \in \mathbb{N}$  we therefore simply need to determine the values of $\wo$ that correspond to roots of
\begin{equation}
\det{\bf B}(\bar{m};\wo) =\det[ {\bf A_0} + \bar{m}^2 {\bf A_1} + \bar{m}^4 {\bf A_2}]=0.
\end{equation} 
In this way we can determine the threshold for instability of each mode number, i.e.~the value $\wo_c^N\left(\bar{m}\right)$ (or $\wo_c^E\left(\bar{m}\right)$ for the everted configuration) at which $\det {\bf B}\left(\bar{m}\right)$ changes sign. The wavenumber that we expect to observe experimentally as the indentation depth is increased is the value of $m$  that minimizes  $\wo_c^N\left(m\right)$ (or $\wo_c^E\left(m\right)$). 

\section{The effect of boundary conditions}\label{APP:NaturalShells}

Previous work by Fitch  \cite{Fitch1968_ijss} showed that clamped shells buckled under indentation, but with a critical mode number that increases as $\lambda$ decreases, apparently in contradiction with our numerical results. However, an important distinction is that the boundary conditions are different in our work (free) and that of Fitch (clamped). In this Appendix, we show that this distinction accounts for the apparent discrepancy. (This comparison is only possible for natural shells because the everted shape is not possible with clamped boundaries.) 

\begin{figure}
	\centering
	\begin{subfigure}{0.49 \linewidth}
		\includegraphics[width=0.95\linewidth]{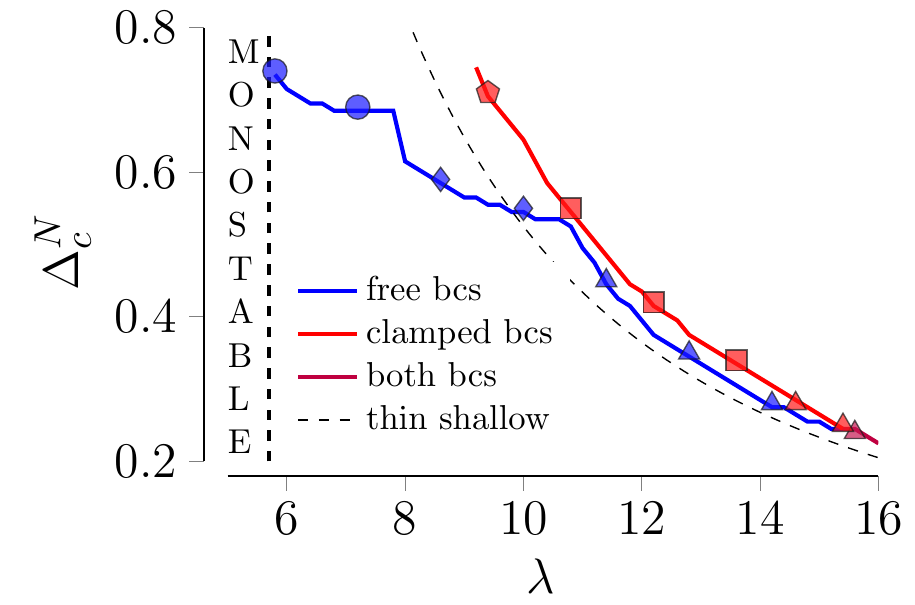}	
		\caption{}
	\end{subfigure}
	\begin{subfigure}{0.49 \linewidth}
		\includegraphics[width=0.95\linewidth]{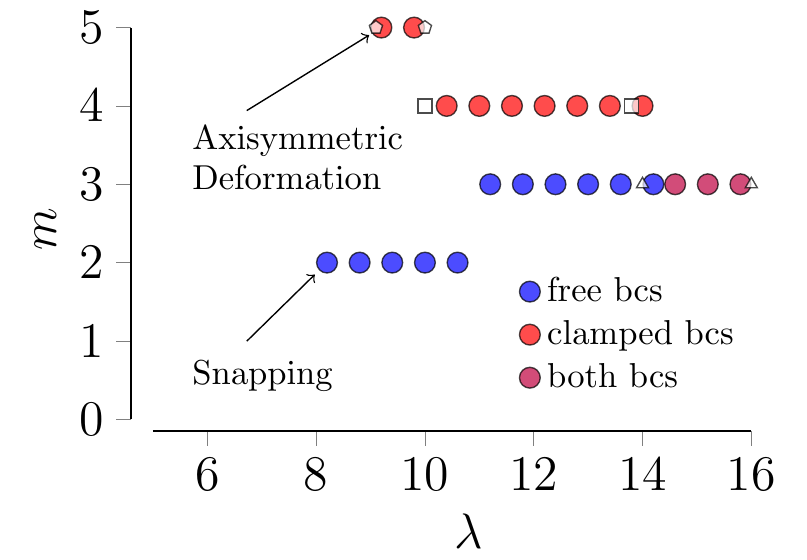}
		\caption{}
	\end{subfigure}
	\caption{Buckling induced by pointwise indentation of a natural shell with either free or clamped boundary conditions. (a) Envelope of the critical indentation depth at the onset of the instability for particular modes, indicated by different markers:  axisymmetric at snap-through,  $\cF=0$ (circles);  $m=2$ (diamonds);  $m=3$ (triangles); $m=4$ (squares);  $m=5$ (pentagons). Blue  indicates modes belonging only to the case of free boundary conditions while red indicates modes that are exclusive to the the clamped case; purple refers to modes that apply to both. The monostable-bistable threshold is also indicated. For large $\lambda$ (thin shells) we expect to recover the result for a thin shallow shell, which corresponds to $\wo\approx 17.5 \left[12\left(1-0.5^2\right)\right]\lambda^{-2}$ and $m=3$, shown as black dashed curve in (a) (this prediction, valid for $\nu=0.5$, has been  obtained analogously to the approach used in \cite{Vella2011_prl} with $\nu=0.3$). (b) Wavenumber of instability at onset for different values of $\lambda$ and either free (filled blue markers) or clamped (filled red markers) boundary conditions. Small white markers (triangles, squares and pentagons) indicate the limiting values of $\lambda$ for which Fitch \cite{Fitch1968_ijss} found that a clamped shell will buckle with 3, 4 and 5 vertices, respectively.} \label{FIG:freeVSclamped}
\end{figure}

We restrict attention to the incompressible case, $\nu=0.5$, for simplicity and to facilitate comparison with our experimental results.
Figure \ref{FIG:freeVSclamped} shows numerical results characterizing  the onset of buckling in indented natural shells with either free or clamped boundary conditions. We observe that for sufficiently thin shells (sufficiently large $\lambda \gtrsim 15$) instability occurs with $m=3$ for both clamped and free boundary conditions; this is consistent with previous numerical results and ABAQUS simulations, which found this triangular mode of instability \cite{Fitch1968_ijss,Vaziri2008_pnas,Vella2011_prl}. However, for relatively thick shells, (sufficiently small $
\lambda$, $5.75\lesssim\lambda\lesssim 15$) we see that the boundary conditions make a qualitative difference to the mode of instability that is observed: for clamped boundary conditions  the mode number of instability increases as $\lambda$ decreases (in agreement with Fitch \cite{Fitch1968_ijss}), while for free boundary conditions, the mode number of instability decreases as $\lambda$ decreases. While this is a numerical result, we rationalize it physically as follows: free shells are able to deform their boundaries to relieve the compression that arises during indentation \cite{Knoche2014_epj,Gomez2016_prs}, while shells with clamped boundaries must remain undisplaced at their edges. We conjecture that this difference in freedom at the edges allows free shells to deform in a prolate--oblate manner, corresponding to the $m=2$ mode. For small values of $\lambda$ there is a lower limit (for both free and clamped conditions) below which the shell does not buckle asymmetrically; this limit is indicated by the two arrows in figure \ref{FIG:freeVSclamped}b. Nevertheless, the response below these two limits are different: a clamped shell with $\lambda \lesssim 9$ exhibits a \emph{mechanical} snapping (i.e.~decrease of applied force but no sudden change of shape), as highlighted by Fitch \cite{Fitch1968_ijss}. However, a free shell with $\lambda \lesssim 8.2$ undergoes a \emph{geometrical} snapping (i.e.~change of shape), before becoming monostable when $\lambda < 5.75$.

Having seen that the effect of free boundary conditions, as opposed to clamped, can be important, we note that the recent work of Sobota \& Seffen \cite{Sobota2017_prsa} investigates the behaviour between these two limiting boundary conditions by examining how the presence of bistability is affected by the stiffnesses of rotational and extensional springs used to connect the shell to the ground. We expect that these intermediate boundary conditions might also have an effect on the robustness of the everted state discussed in this paper, but leave a discussion of such an effect to future work.

\section{Wall model}\label{APP:wall}
The  analysis presented in the main text makes use of a simply supported boundary condition on the shell. This produces results that are qualitatively in agreement with experiments, but with some disagreement.
In this Section we discuss a `\emph{wall model}', which is intended to capture  the interaction between the shell and a planar rigid substrate (as in our experiments): rather than having freely supported edges (as supposed in the calculations presented in the main text), the contact point between the shell and the substrate may move during the course of indentation --- this contact is not necessarily located at the outer edge. Our aim in including this complication is to improve the quantitative agreement between theory and experiment.

To include the effect of a planar supporting boundary in our shallow shell model, we consider an impenetrable wall, which is incorporated using an energy density of the form
\begin{equation}
E_{\mathrm{wall}} = \mathcal{P}\, \mathcal{U}_{H} \left(Z_{\mathrm{wall}} - Z\left(\rho\right)\right) \mathcal{U}_{H}\left(\rho - \bar{\rho}\right).
\label{eqn:Lagrange}
\end{equation}
Here we use a logistic function $\mathcal{U}_{H}(x) = (1+e^{-kx})^{-1}$ with a large value of $k$ to ensure that penetration of the wall is strongly penalized (first appearance of $U_H(x)$). The second appearance of $\mathcal{U}_{H}(x)$  limits the radial extent of the \emph{wall} to $\rho \ge \bar{\rho}=0.5$; this threshold is arbitrarily chosen to allow the apex of the shell to cross the wall while allowing the contact point to move radially as expected by the experiments. A sufficiently large $k$ should suffice to ensure these features occur only in the desired region and do not unduly influence the behaviour elsewhere;  in practice we find that $k=500$ gives a good trade-off between having a perfectly impenetrable wall, (i.e.~the logistic function tends to the Heaviside step function as $k \rightarrow \infty$), and guaranteeing convergence of the numerical code. In \eqref{eqn:Lagrange},  $\mathcal{P}$ is a Lagrange multiplier that is associated with the contact force and $Z=Z_{\mathrm{wall}}$ is the position of the substrate/wall.
The energy $E_{\mathrm{wall}}$ is added to the elastic energies ($E_{\mathrm{stretching}}+E_{\mathrm{bending}}$, which are discussed, for example, by  Ventsel and Krauthammer \cite{Ventsel2001}), to obtain the Lagrangian  $\mathcal{L} = E_{\mathrm{bending}}+E_{\mathrm{stretching}} +E_{\mathrm{wall}}$.  Minimizing this energy with respect to the unknown displacement of the shell and the Lagrange multiplier, we obtain a new system of equations that describes the equilibrium of the shell. In the axisymmetric framework it becomes
\begin{equation}
\begin{cases}
\left(\frac{1}{\lambda_s}\right)^{4}\rho \frac{\upd }{\upd \rho} \left(\nabla^2 W\right) + \frac{\upd \Phi}{\upd \rho} \left(\rho - \frac{\upd W}{\upd \rho}\right) + \mathcal{P}\, \mathcal{U}_{\delta} \left(Z_{\mathrm{wall}} - Z\left(\rho\right)\right)\mathcal{U}_{H} \left(\rho -\bar{\rho}\right) = \frac{\mathcal{F}}{2 \pi} \\
\rho \frac{\upd}{\upd \rho}\left(\frac{1}{\rho}\frac{\upd}{\upd \rho} \left(\rho \frac{\upd \Phi}{\upd \rho} \right)\right)= \rho \frac{\upd W}{\upd \rho} - \frac{1}{2}\left(\frac{\upd W}{\upd \rho}\right)^2 \\
\mathcal{U}_{H} \left(Z_{\mathrm{wall}} - Z\left(\rho\right)\right)\mathcal{U}_{H} \left(\rho - \bar{\rho}\right) = 0
\end{cases}
\end{equation}
where $\mathcal{U}_{\delta}\left(x\right) = \upd \mathcal{U}_{H}(x) / \upd x = k e^{-kx}(1+e^{-kx})^{-2}$. The boundary conditions in table \ref{TAB:BCs_zeroth} for the free case are modified, as detailed in  table \ref{TAB:BCs_zeroth_wall}. 

\begin{table}[h]
	\begin{center}
		\caption{The seven boundary conditions required to solve the zeroth order axisimmetrical problem of the \emph{wall} model.		\label{TAB:BCs_zeroth_wall}}
		\begin{tabular}{p{0.09\linewidth} p{0.27\linewidth} p{0.54\linewidth}}
			& & \\ 
			\hline
			\multicolumn{3}{|c|}{Boundary Conditions for the zeroth order \emph{wall} model } \\
			\hline
			\multirow{1}{*}{$\rho=0$} & Zero normal displacement & $W^{(0)}(0)=0$ \\ 
			\multirow{1}{*}{$\rho=0$} & Zero radial displacement & $U_r^{(0)}(0)=\lim\limits_{\rho \rightarrow 0}\left[\rho \frac{\upd^2 \Phi^{(0)}}{\upd \rho^2} - \nu \frac{\upd \Phi^{(0)}}{\upd \rho} \right] =0$ \\ 
			\multirow{1}{*}{$\rho=0$} & Avoid cusp & $\frac{\upd W^{(0)}}{\upd \rho}\bigg\rvert_{\rho=0}=0$ \\ 
			\multirow{1}{*}{$\rho=0$} & No wall & $\mathcal{P}^{(0)}(0) =0$ \\ 
			\multirow{1}{*}{$\rho=1$} & Zero radial stress & $\Srr^{(0)}(1) = \left[\frac{1}{\rho}\frac{\upd \Phi^{(0)}}{\upd \rho}\right]\bigg\rvert_{\rho=1}=0$ \\ 
			\multirow{1}{*}{$\rho=1$} & Zero bending & $M_r^{(0)}(1)=\left[\frac{\upd^2 W^{(0)}}{\upd\rho^2} + \frac{\nu}{\rho} \frac{\upd W^{(0)}}{\upd\rho}\right]\bigg\rvert_{\rho=1}=0$ \\ 
			\multirow{1}{*}{$\rho=1$} & Zero shear force & $Q_r^{(0)}(1)=\left[\frac{\upd^3 W^{(0)}}{\upd\rho^3} + \frac{1}{\rho} \frac{\upd^2 W^{(0)}}{\upd\rho^2} - \frac{1}{\rho^2} \frac{\upd W^{(0)}}{\upd\rho}\right]\bigg\rvert_{\rho=1} = 0$ \\ 
			\hline
		\end{tabular}
	\end{center}
\end{table}

\begin{figure}[h]
	\centering
	\begin{subfigure}{0.46 \linewidth}
		\includegraphics[width=0.95\linewidth]{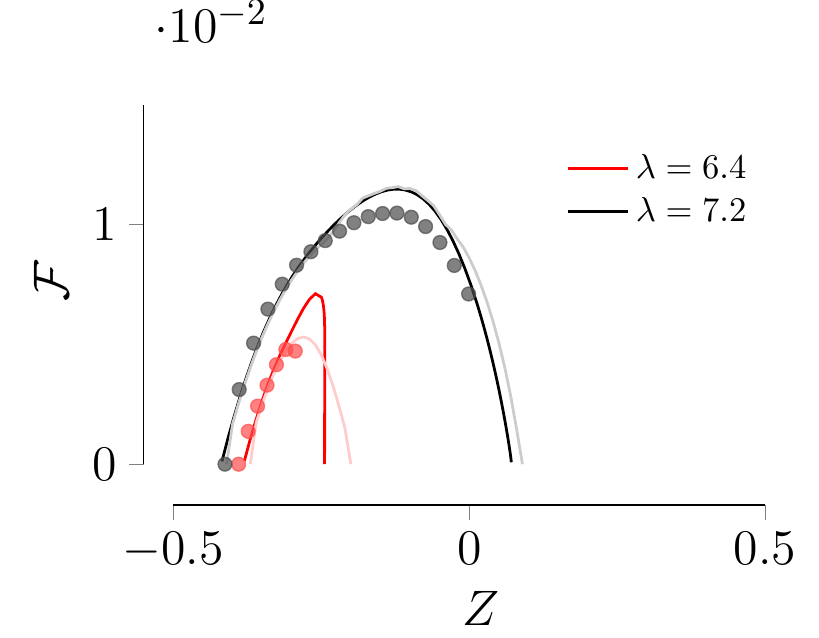}	
		\caption{}
	\end{subfigure}
	\begin{subfigure}{0.51 \linewidth}
		\includegraphics[width=0.95\linewidth]{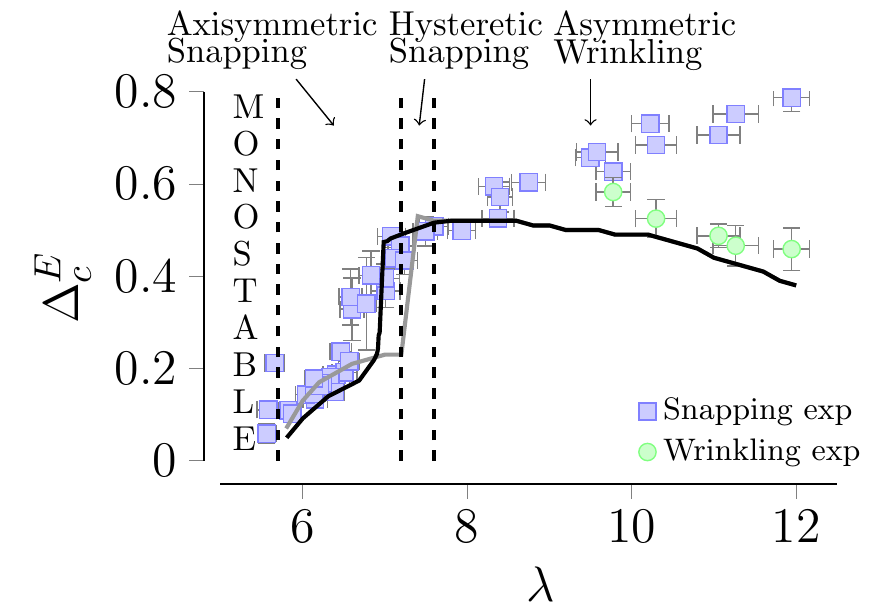}
		\caption{}
	\end{subfigure}
	\caption{(a) Force indentation curve from the two everted configurations obtained using $\lambda=6.4$ (red) and $\lambda=7.2$ (black) and employing the \emph{wall} model. Experimental data are superimposed. Lighter colour curves represent the envelope of the continuous curves in figure \ref{FIG:SYMvsASYM_snapping} of the main text (obtained with the model used in the main text). (b) Regime diagram for the indentation from the everted shape obtained using the \emph{wall} model (black curve). As a comparison, the model described in the main text is highlighted in grey. Experimental data (markers).}
	\label{FIG:FD_wall}       
\end{figure}

This more detailed model is able to explain two of the intricate features of the physics of the system that were highlighted by the experiments. Firstly, the experimental force--indentation curve exhibits a more sudden drop after the force peak (red dots in figure \ref{FIG:SYMvsASYM_snapping}a) than is expected based on the simple free boundary conditions. Figure \ref{FIG:FD_wall}(a) shows the force--displacement curves predicted by this new model, together with the experimental data and the original simulation results. We see that the effect of the substrate/wall is to make the drop-off in force after the peak much more sudden. To understand this, we note that the everted shape is characterized by a boundary layer around the outer boundary whose width  scales as $\lambda^{-1}$. For smaller $\lambda$,  there is thus a larger and larger region that is almost flat or that has a curvature of the same sign as the natural configuration; while indenting, this region forces the overall shell to snapback more easily onto the natural configuration than in the case of a normally constrained outer edge. In the force--displacement plot, the zone in which $\mathcal{F}$ decreases therefore becomes steeper and more difficult to detect experimentally. We note that the effect of the substrate becomes less evident as  $\lambda$ increases (consistent with the above explanation since the extent of the boundary layer decreases as $\lambda$ increases).

A second feature of the experiments is the sharp increase in the critical indentation depth in the range of $\lambda$'s highlighted as hysteretic snapping in figure \ref{FIG:FD_wall}(b): while a similar increase was observed in the numerical results of the main text, the range of $\lambda$ over which it occurred did not correspond to that observed experimentally. Figure  \ref{FIG:FD_wall}(b) shows that with this modified boundary condition, the increase occurs at a significantly smaller value of $\lambda$, closer to that observed experimentally.

\end{document}